\documentclass[11pt,a4paper]{article}
\usepackage[utf8]{inputenc}
\usepackage{a4wide}
\usepackage{amsmath,amssymb}
\usepackage{amsfonts}
\usepackage{cite}
\usepackage{url}
\usepackage{color}
\usepackage{xcolor}
\usepackage{adjustbox}
\usepackage{multirow}
\usepackage{pdflscape}
\usepackage{textcomp}
\usepackage{gensymb}
\usepackage{graphicx}
\usepackage{diagbox}
\usepackage{pifont}
\usepackage{float}
\usepackage{bigstrut}
\usepackage[normalem]{ulem}
\usepackage[small,bf]{caption}
\usepackage{subcaption}
\usepackage{tabulary}
\usepackage{booktabs}
\usepackage{stackengine}
\usepackage{mathtools}
\usepackage{enumerate}
\usepackage{makecell}
\usepackage{listings}
\usepackage{bbm}
\usepackage[unicode]{hyperref}
\usepackage{placeins}
\usepackage{cleveref}
\usepackage{ragged2e}
\usepackage{upgreek}
\usepackage{tabularx}
\definecolor{greenLinks}{rgb}{0, 0.6, 0} 
\definecolor{blueLinks}{rgb}{0, 0, 0.6}
\definecolor{redLinks}{rgb}{0.6, 0, 0}
\definecolor{eprintLinks}{rgb}{0.4, 0.4, 0.4}
\definecolor{journalLinks}{rgb}{0.6, 0, 0}
\begin{document}

\pagenumbering{Alph}
\begin{titlepage}

\vspace*{15mm}

\begin{center}
{ \bf\LARGE {Testable neutrino mass and TeV-scale leptogenesis \\[2mm] in a \texorpdfstring{$D_4$}{D4} inverse seesaw model}}\\[8mm]
Qiu Yan$^{\,a,}$,
Yakefu Reyimuaji$^{\,a,}$\footnote{E-mail: \texttt{yreyi@hotmail.com}},
Xianshuo Zhang$^{\,a,}$\\

\vspace{8mm}
$^{a}$\,{\it School of Physical Science and Technology, Xinjiang University, Urumqi, Xinjiang 830017, China} \\
\vspace{2mm}

\end{center}
\setcounter{footnote}{0} 

\vspace{8mm}

\begin{abstract}
\noindent An inverse seesaw model for neutrino masses and mixing is proposed, based on the spontaneous breaking of a $D_4$ flavor symmetry. The model simultaneously accounts for the neutrino oscillation parameters, the baryon asymmetry of the Universe through TeV-scale leptogenesis, and makes predictions for charged-lepton flavor violation (cLFV) that are consistent with present experimental limits. A phenomenological analysis shows that the model is consistent with the observed neutrino oscillation data for both normal and inverted mass orderings. Successful leptogenesis is realized for the lightest pseudo-Dirac pair in the multi-TeV range. The predicted cLFV branching ratios lie well below experimental upper bounds, ensuring compatibility with existing constraints. These results establish the model as a viable framework that links low-energy neutrino observables to TeV-scale leptogenesis and cLFV phenomenology.

\end{abstract}

\end{titlepage}
\pagenumbering{arabic}



\section{Introduction}
\label{sec:intro}

The Standard Model (SM) of particle physics is one of the most successful theoretical frameworks in modern physics, providing an accurate description of fundamental particles and their interactions. Despite its remarkable achievements, the SM falls short of being a complete theory. It fails to account for several crucial experimental observations. Among these, neutrino oscillation experiments have firmly established that neutrinos possess tiny but nonzero masses and exhibit nontrivial mixing~\cite{Esteban:2020cvm}. At the same time, cosmological observations have determined the baryon-to-entropy ratio of the Universe to be $Y_B \simeq (8.718 \pm 0.047)\times 10^{-11}$~\cite{Planck:2018vyg}, a value that exceeds by several orders of magnitude what can be generated within the SM. These two long-standing puzzles--the origin of neutrino masses and the generation of matter-antimatter asymmetry--point to the need for new physics beyond the SM.

Considerable progress has been made toward addressing these two fundamental issues. A classic and elegant solution is the thermal leptogenesis scenario~\cite{Fukugita:1986hr,Minkowski:1977sc,Mohapatra:1980yp}, which is naturally embedded in the conventional type-I seesaw mechanism.  In this framework, the observed baryon asymmetry is generated through the out-of-equilibrium CP-violating decays of heavy right-handed neutrinos. However, to simultaneously reproduce the observed neutrino masses and the baryon asymmetry, heavy neutrinos are typically required to have masses as high as $ 10^9\ \mathrm{GeV}$~\cite{Davidson:2002qv,Davidson:2008bu}. This limitation motivates the construction of alternative models that can reduce the seesaw scale.

The inverse seesaw mechanism~\cite{Mohapatra:1986bd,Kang:2006sn,Hirsch:2006ft,Romao:2007jr,Zhang:2025dsa} offers a particularly attractive pathway. By introducing pairs of nearly degenerate sterile neutrinos, it naturally accommodates TeV-scale heavy neutrinos without requiring unnaturally small Yukawa couplings. Consequently, inverse seesaw models are endowed with a rich set of phenomenological predictions that can be directly tested in current and future collider experiments~\cite{Das:2012ze,Das:2017nvm,Das:2016hof,Bhardwaj:2018lma,Das:2018usr,ThomasArun:2021rwf,Mondal:2016kof}.

Within the inverse seesaw (ISS) framework, the simplest viable realization of leptogenesis is the minimal ISS(2,2) model. Here,  ``ISS(2,2)" denotes the inverse seesaw model with two right-handed neutrinos and two additional singlet fermions. This model extends the SM by these four fermionic singlets. The heavy neutrinos in this setup form nearly degenerate pseudo-Dirac pairs. Their CP-violating decays can be resonantly enhanced through the interference between tree-level and one-loop amplitudes. This resonant enhancement generates a sizable lepton asymmetry at the TeV scale~\cite {Pilaftsis:2005rv,Flanz:1994yx,Pilaftsis:2003gt,Cirigliano:2006nu,Zhang:2026kyy}. The asymmetry is subsequently converted into the baryon asymmetry via electroweak sphaleron processes~\cite {Klinkhamer:1984di,Arnold:1987mh}. The ISS mechanism thus provides a natural bridge between low-scale neutrino mass generation and testable leptogenesis.

In this work, we construct an ISS(2,2) model based on the $D_4$ symmetry. The model simultaneously reproduces the neutrino masses and mixing, consistent with the global fit to oscillation data. It realizes the thermal leptogenesis at the TeV scale. Previous studies of $D_4$ symmetry were mostly confined to the high-scale regime~\cite{Vien:2014ica,Vien:2020uzf,Vien:2021diw,Miskaoui:2021nbm,Vien:2023xyq} and did not address leptogenesis. Low-scale inverse seesaw realizations with $D_4$ symmetry have also been explored~\cite{Hernandez:2021mxo}. There,  $D_4$ is supplemented by a direct product of cyclic groups, $Z_4 \times Z_3^{(1)} \times Z_3^{(2)} \times Z_{16}$. Twelve singlet scalars are introduced to reproduce the fermion masses and mixing. Leptogenesis, however, is not addressed. By contrast, the present work unifies neutrino masses and mixing with TeV-scale thermal leptogenesis in a more economical setup, requiring only a $D_4$ doublet scalar and a singlet scalar beyond the ISS(2,2) fermion content. We also investigate cLFV. The branching ratios of $\mu\to e\gamma$, $\tau\to e\gamma$, and $\tau\to\mu\gamma$ are calculated. The numerical analysis shows that, within the parameter space that simultaneously satisfies the neutrino oscillation data and the requirements of thermal leptogenesis, the predicted branching ratios lie many orders of magnitude below current experimental limits set by MEG II~\cite{MEGII:2023ltw}, BaBar~\cite{BaBar:2009hkt}, and Belle~\cite{Belle:2021ysv}. 

The organization of this paper is as follows. In section~\ref{sec:model}, the theoretical framework is introduced, including the field content, symmetry assignments, and the derived mass matrices along with the corresponding observables. Section~\ref{sec:pheno} is devoted to a phenomenological analysis, where neutrino oscillations, thermal leptogenesis, and cLFV processes are studied. In section~\ref{sec:conclusions}, the results are summarized and possible future directions are outlined. The relevant properties of the $D_4$ group and the forbidden operators are presented in appendices~\ref{app:sol4polyeq} and~\ref{app:DZ5forb}. The results of thermal leptogenesis  for inverted ordering are provided in appendix~\ref{app:IO}.

\section{The model}
\label{sec:model}

\subsection{ Field content and symmetries}
\label{subsec:fldsymtry}

The model is based on the SM gauge symmetry $SU(2)_L \times U(1)_Y$, extended by a flavor symmetry $D_4$ and an auxiliary discrete symmetry $Z_5$. These extended symmetries are essential for forbidding undesired mass terms, as detailed in appendix~\ref{app:DZ5forb}.
Neutrino masses are generated via the ISS mechanism. In addition to the SM field content, the model introduces a $D_4$ doublet of right-handed neutrinos $N_D$, a gauge singlet but $D_4$ doublet fermions $\chi_D$, a scalar doublet $\eta$, and a scalar singlet $S$. These additional fields constitute the minimal particle content required for the ISS(2,2) realization. The $D_4$ symmetry is spontaneously broken by the vacuum expectation value (VEV) of the $\eta$ field, while the $Z_5$ symmetry is broken by the VEVs $\langle S \rangle$ and $\langle H \rangle$ of the $S$ and $H$ fields, respectively.

Under the full symmetry group, the charged-lepton mass matrix $M_{\mathrm{CL}}$ and the neutrino mass matrix $m_\nu$ can be decomposed as
\begin{equation}
	\begin{aligned}
		M_{\mathrm{CL}} &= M_{\mathrm{CL}}^{(0)}+M_{\mathrm{CL}}^{(1)},\\
		m_{\nu} &=m_{\nu}^{(0)}+m_{\nu}^{(1)}.
	\end{aligned}
	\label{eq:massdecomp}
\end{equation}
The terms $M_{\mathrm{CL}}^{(0)}$ and $m_{\nu}^{(0)}$ remain invariant under $D_4$ transformations and are therefore present in the unbroken flavor-symmetry limit. By contrast, $M_{\mathrm{CL}}^{(1)}$ and $m_{\nu}^{(1)}$ arise from the spontaneous breaking of the $D_4$ symmetry when the scalar field $\eta$ acquires a VEV, and consequently vanish in the symmetry limit. The representation assignments of all fields under the gauge and flavor symmetries are listed in table~\ref{tab:fldcntchrgasgn}.

\begin{table}[h]
    \centering
    \aboverulesep=0pt
    \belowrulesep=0pt
		 \begin{tabular}{|c||c|c|c|c|c|c||c|c|c|}
			\toprule 
          \text{Fields}&$ L_{e}$ & $ L_{D}$&$ e_{D}$ &$ e_R$ & $ N_{D}$ & $\chi_D$&$ \eta$&$S$ &$H$\\
			\midrule
			$SU(2)_{L}$ & $2$ & $2$ &$1$&$1$&$1$ & $1$&$1$&$1$&$2$\\
			$U(1)_{Y}$ &-$\frac{1}{2}$&-$\frac{1}{2}$&$-1$&$-1$& 0& 0&0&0&$\frac{1}{2}$\\
			$D_4$ & $1_{+-}$& $2$&$2$&$1_{+-}$& $2$ & $2$&$2$&$1_{++}$&$1_{++}$\\
            $Z_{5}$&$q^4$&$q^4$&$q$&$q$&$q^2$&$q^3$&$1$&$q^2$&$q^3$ \\
			\bottomrule
		\end{tabular}
		\caption{Field content of the model, and their transformation properties under the gauge and flavor symmetries.}
        \label{tab:fldcntchrgasgn}
\end{table}

Based on the symmetries specified in table~\ref{tab:fldcntchrgasgn}, the most general invariant Lagrangian can be written as the sum of two parts:
\begin{equation}
	\label{eq:totlag}
	\mathcal{L} = \mathcal{L}_0 + \mathcal{L}_1,
\end{equation}
where $\mathcal{L}_0$ contains the renormalizable terms that are invariant under the full symmetry group, while $\mathcal{L}_1$ consists of higher-dimensional operators that break the $D_4$ symmetry spontaneously after $\eta$ acquires a VEV. All terms in $\mathcal{L}$ respect the $SU(2)_L \times U(1)_Y$ gauge symmetry and $D_4 \times Z_5$ flavor symmetry.

The renormalizable part of the Lagrangian is given by
\begin{equation}
	\begin{aligned}
		\mathcal{L}_0 =&\, \alpha \overline{L_e} H e_R + \beta \overline{L_D} H e_D + \alpha_D \overline{L_D} \tilde{H} N_D + \mu_D \overline{N_D^c }\chi_D +\ \mathrm{H.c.}.
	\end{aligned}
	\label{eq:lagrangian0}
\end{equation}
The first two terms generate the charged lepton masses through couplings to the SM Higgs doublet $H$. The third term provides the Dirac mass connecting the left-handed lepton doublets to the right-handed neutrinos. The fourth term is a gauge-invariant mass term that mixes the $N_D$ and $\chi_D$ fields.

The higher-dimensional operators that break the $D_4$ symmetry spontaneously are contained in $\mathcal{L}_1$. These operators are suppressed by the cutoff scale $\Lambda$ of the flavor model, and become effective after the scalar field $\eta$ acquires a VEV. The full expression for $\mathcal{L}_1$ is given by
\begin{equation}
	\begin{aligned}
		\mathcal{L}_1 = \frac{1}{\Lambda} \biggl(
		& k_1\overline{L_e} H e_D \eta + k_2\overline{L_D} H e_R \eta + k_3 \overline{L_e} \tilde{H} N_D \eta \\
		& + k_{4a} \overline{N_D^c} \chi_D (\eta^2)_{1_{++}} + k_{4b} \overline{N_D^c} \chi_D (\eta^2)_{1_{+-}} + k_{4c} \overline{N_D^c} \chi_D (\eta^2)_{1_{-+}} \\
		& + k_5\overline{N_D^c} \chi_D (H^\dagger H) + k_6\overline{N_D^c} \chi_D (S^* S) \\
		& + k_7 \overline{\chi_D^c} \chi_D S^2 + k_8\overline{N_D^c} N_D (S^*)^2 \biggr) + \mathrm{H.c.}.
	\end{aligned}
	\label{eq:lagrangian1}
\end{equation}
Here $L_D$ denotes the $D_4$ doublet of left-handed charged lepton fields, defined as $L_D=(L_{\mu},L_{\tau})^T$, and $e_D$ denotes the corresponding $D_4$ doublet of right-handed charged leptons, $e_D=(\mu_R,\tau_R)^T$. The first two terms in $\mathcal{L}_1$ involve couplings of the lepton fields to $\eta$, and are responsible for generating the off-diagonal entries in the charged-lepton mass matrix after $\eta$ acquires a VEV. The remaining terms contribute to the neutrino mass matrix and the mass splittings among the sterile neutrinos.

\subsection{ Yukawa sector and lepton masses}
\label{subsec:yukfermass}

With the VEV alignment of the Higgs $\langle H\rangle=(0,v/\sqrt{2})^T$ as well as the scalar $\langle\eta\rangle=(v_{\eta}/\sqrt{2},0)^T$~\footnote{There exist multiple VEV alignments  $\langle\eta\rangle=(v_{1},v_{2})^T/\sqrt{2}$ for the scalar doublet $\eta$ transforming under the $D_4$ group: $v_1=v_\eta,\;v_2=0$; $v_1=0,\;v_2=v_\eta$;  $v_1=v_2=v_\eta$, $v_1 \neq v_2$ and both nonzero~\cite{Vien:2013zra,deMedeirosVarzielas:2025byb}. An analysis of the scalar potential shows that the potential reaches its global minimum when the VEVs take the form $(v_\eta, 0)$ or $(0, v_\eta)$. In contrast, the configurations $(v_\eta, v_\eta)$ and $v_1 \neq v_2$ correspond to relatively higher potential energies at the minima. This conclusion requires the model's coupling constants to satisfy $\lambda_{\eta,1} + \lambda_{\eta,2} > 0$. Therefore, without loss of generality, we choose the first configuration $(v_\eta, 0)$ as the physical vacuum. }, the charged-lepton mass matrix can be derived from the terms in $\mathcal{L}_0$ and $\mathcal{L}_1$. It takes the form
\begin{equation}
	M_{\mathrm{CL}}=
	\frac{v}{\sqrt{2} } \begin{pmatrix}
		\alpha  & k_1\frac{v_{\eta}}{\Lambda} & 0\\
		k_2\frac{v_{\eta}}{\Lambda} & 0 & \beta \\
		0 & \beta  &0
	\end{pmatrix}.
	\label{eq:clmass}
\end{equation}
The entries proportional to $\alpha v$ and $\beta v$ are generated by the renormalizable couplings in $\mathcal{L}_0$, while the off-diagonal entries arise from the higher-dimensional operators in $\mathcal{L}_1$. The specific texture of this mass matrix, with zeros in the main off-diagonal  positions, is a consequence of the $D_4$ and $Z_5$ charge assignments listed in Table~\ref{tab:fldcntchrgasgn}.

The charged-lepton mass matrix $M_{\mathrm{CL}}$ is diagonalized by a unitary transformation
$U_l^\dagger M_{\text{CL}}$ $ M_{\text{CL}}^\dagger U_l
= \mathrm{diag}(m_e^2,\,m_\mu^2,\,m_\tau^2)$. The eigenvalues $m_l^2$, with $l= e, \mu, \tau$, are given by
\begin{equation}
	\label{eq:CE}
	\begin{aligned}
		m_l^2 &= \frac{T_1}{3} + 2\sqrt{-\frac{p}{3}}\;
		\cos\!\left[\frac{1}{3}\arccos\!\left(\frac{3q}{2p}\sqrt{-\frac{3}{p}}\right) - \frac{2\pi}{3}\,n_l \right],
	\end{aligned}
\end{equation}
where  $n_e=2$, $n_\mu=1$, $n_\tau=0$, and 
\begin{equation}
	\begin{aligned}
		T_1 &= \frac{v^2}{2}\bigl(|\alpha|^2+2|\beta|^2\bigr)
		+ \frac{v^2v_\eta^2}{2\Lambda^2}\bigl(|k_1|^2+|k_2|^2\bigr),\\[4pt]
		T_2 &= \frac{v^4}{4}|\beta|^2\bigl(2|\alpha|^2+|\beta|^2\bigr)
		+ \frac{v^4 v_\eta^2}{4\Lambda^2}|\beta|^2\bigl(|k_1|^2+|k_2|^2\bigr)
		+ \frac{v^4v_\eta^4}{4\Lambda^4}|k_1|^2|k_2|^2,\\[4pt]
		T_3 &= \frac{v^6}{8}|\alpha|^2|\beta|^4,\\
		p &= T_2 - \frac{T_1^2}{3},\\
		q &= \frac{1}{27} \left(9T_1T_2-2T_1^3\right) - T_3,
	\end{aligned}
\end{equation}
satisfying the conditions $p <  0$ and $4p^3 +27q^2<  0$ to ensure three positive eigenvalues. The corresponding normalized eigenvectors form the columns of the unitary matrix $U_l = (U_e,\,U_{\mu},\,U_{\tau})$, with
\begin{equation}
	\begin{gathered}
		U_l = \frac{1}{\mathcal{N}_l}
		\begin{pmatrix}
			\left[ m_l^2-\frac{1}{2}|\beta|^2 v^2 - |k_2|^2\frac{v^2v_{\eta}^2}{2\Lambda^2}\right]\left[m_l^2-\frac{1}{2}|\beta|^2 v^2\right] \\[8pt]
			\frac{v ^2v_{\eta}}{2\Lambda} \alpha^* k_2\left[m_l^2-\frac{1}{2}|\beta|^2 v^2\right]  \\[4pt]
			\frac{v^2 v_{\eta}}{2\Lambda} \beta k_1^*\left[ m_l^2-\frac{1}{2}|\beta|^2 v^2 - |k_2|^2\frac{v^2v_{\eta}^2}{2\Lambda^2}\right]
		\end{pmatrix}, 
	\end{gathered}
	\label{eq:clmm}
\end{equation}
where $\mathcal{N}_l$ is the normalization factor, defined as the square root of the sum of the squared moduli of the three entries in eq.~\eqref{eq:clmm}.

Turning now to the neutrino sector, the fields $N_D\equiv (N_1,N_2)\sim 2$ and $\chi_D\equiv (\chi_1,\chi_2)\sim 2$ are assigned as doublets under $D_4$. From the Lagrangian $\mathcal{L}$, the mass matrices for the neutral fermions take the following forms: 
\begin{equation}
\mu = \mu_0 \begin{pmatrix} 0 & 1 \\ 1 & 0 \end{pmatrix},\qquad
M_D = \begin{pmatrix} y_1 & 0 \\ 0 & Y_1 \\ Y_1 & 0 \end{pmatrix},\qquad
M_{R} = \begin{pmatrix} M_1 & M \\ M & M_2 \end{pmatrix},
\label{eq:masmatinv}
\end{equation}
where the parameters are defined as
\begin{equation}
\begin{aligned}
y_1 &= k_3\frac{v_{\eta}}{2\Lambda}v, & Y_1 &= \frac{\alpha_D v}{\sqrt{2}},\\
M   &= k_5\frac{v^2}{2\Lambda}+k_6\frac{u^2}{2\Lambda}+\mu_D, & \mu_0 &= k_7\frac{u^2}{2\Lambda},\\
M_1 &= (k_{4b}+k_{4c})\frac{v_{\eta}^2}{2\Lambda}, & M_2 &= (k_{4b}-k_{4c})\frac{v_{\eta}^2}{2\Lambda},
\end{aligned}
\end{equation}
The matrix $\mu$ is the Majorana mass matrix for the singlet fermions $\chi_D$, while $M_D$ is the Dirac mass matrix connecting the left-handed neutrinos to the right-handed neutrinos. The matrix $M_R$ is the $N_D-\chi_D$ mixing mass matrix. The VEVs $u$ and $v_\eta$ appear in the expansion of the scalars
\begin{equation}
	S = \frac{1}{\sqrt{2}} (u+\phi_S) e^{i\rho}, \quad
	\eta = \frac{1}{\sqrt{2}}\begin{pmatrix}
		v_{\eta}+\eta_1\\
		\eta_2
	\end{pmatrix}.
	\label{eq:sclrvevexp}
\end{equation}
The effective light neutrino mass matrix is then generated through the ISS mechanism,
\begin{equation}
\begin{aligned}
m_{\nu} &= M_D\,(M_{\mathrm{R}}^T)^{-1}\,\mu\,M_{\mathrm{R}}^{-1}\,M_D^T.
\end{aligned}
\end{equation}
Details of the generic block-diagonalization procedure used to obtain this result can be found, e.g., in Ref.~\cite{Wang:2024qhe}. A common feature of the ISS construction is the mass hierarchy $O(M_{\mathrm{R}})\gg O(M_D)\gg O(\mu)$. This hierarchy is realized through the VEV  hierarchy of the scalar fields $\langle\eta\rangle \gg \langle S\rangle \sim \langle H\rangle$. Consequently, the 
$D_4$ symmetry is broken at a higher scale, while the 
$Z_5$ and electroweak symmetries are broken at a lower scale. This choice also forbids unwanted decays of the Higgs boson. The smallness of $\mu$ originates entirely from the suppressed dimension‑five operators in $\mathcal{L}_1$. With the mass matrices in eq.~\eqref{eq:masmatinv} and the ISS mechanism above, the effective light neutrino mass matrix can be computed explicitly. In the unbroken limit of $D_4$, the invariant contribution is given by
 \begin{align}
   m_{\nu}^{(0)}=  \frac{ Y_1^2 \mu_0}{M^2} \begin{pmatrix}
 0 &0  &0 \\
  0&  0&1 \\
  0& 1 &0
\end{pmatrix}.
\label{eq:nu0}
 \end{align}
This matrix has a characteristic texture with zeros in the first row and column, which is a consequence of the extended symmetries.  When the scalar field $\eta$ acquires a VEV, the symmetry is spontaneously broken, and finally the light neutrino mass matrix can be written as
\begin{align}
    m_{\nu}=\frac{\mu_0}{(M^2-M_1M_2)^2} \begin{pmatrix}
-2MM_2y_1^2  & (M_1M_2+M^2)Y_1y_1 & -2MM_2Y_1y_1 \\
 (M_1M_2+M^2)Y_1y_1 & -2MM_1Y_1^2 & (M_1M_2+M^2)Y_1^2 \\
 -2MM_2Y_1y_1 & (M_1M_2+M^2)Y_1^2 & -2MM_2Y_1^2
\end{pmatrix}.
\end{align}
This general texture, with all entries filled after symmetry breaking, provides sufficient freedom to accommodate the observed neutrino mixing. The structure of the mass matrix yields one massless neutrino for both normal and inverted orderings. This is a consequence of the minimal particle content of the ISS(2,2) framework, which contains only two heavy sterile neutrino pairs. It is also apparent from the matrix structure, where the first and third raws (or columns) are proportional to each other. This reduces rank of the matrix.

For complex Yukawa couplings, the light neutrino masses are obtained by diagonalization $U^T_\nu m_\nu U_\nu =\mathrm{diag}(0,\,m_-,\,m_+)$, with $m_2=m_-, m_3=m_+$ for normal ordering (NO) while $m_1=m_-, m_2=m_+$ for inverted ordering (IO). The nonzero neutrino mass squares are expressed as
\begin{equation}
		m_{\mp}^2  = \dfrac{s_1 \mp \sqrt{s_1^{2}-4s_2}}{2}, 
\end{equation}
where
\begin{equation}
	\begin{aligned}
		s_1&=\sum_{i,j} \left| \left(m_\nu\right)_{ij} \right|^2,\\
		s_2&=\sum_{i<j}\sum_{k<l}
		\left|(m_\nu)_{ik}(m_\nu)_{jl}-(m_\nu)_{il}(m_\nu)_{jk}\right|^2,
	\end{aligned}
\end{equation}
which are basis-independent constraints~\cite{Reyimuaji:2024kqs}.

The unitary matrix $U_\nu$ is given by $U_\nu = (u_0,u_-,u_+)$ for NO and by $U_\nu = (u_-,u_+,u_0)$ for IO. Their components are 
\begin{equation}
\begin{aligned}
(u_0)_i & = \frac{1}{\mathcal{N}_0} \sum_{j,k} \epsilon_{ijk} \,(m_\nu)_{1j}^*\,(m_\nu)_{2k}^*, \\[6pt]
(u_\pm)_i & = \frac{1}{\mathcal{N}_\pm}
\sum_{j,k,l,m} \epsilon_{3jk}\,\epsilon_{ilm}
\left(
\sum_p (m_\nu)_{jp}\; (m_\nu)_{lp}^* - m_\pm \delta_{jl}
\right)
\left(
\sum_q (m_\nu)_{kq}\;(m_\nu)_{mq}^* - m_\pm \delta_{km}
\right),
\end{aligned}
\end{equation}
where $\mathcal{N}_0$ and $\mathcal{N}_{\pm}$ are normalization constants, given by
\begin{equation}
    \begin{aligned}
        \mathcal{N}_0^2 & = \sum_{i=1}^3 \left| \sum_{j,k=1}^3 \epsilon_{ijk} \;(m_\nu)_{1j}^*\,(m_\nu)_{2k}^* \right|^2,\\
        \mathcal{N}_\pm^2 &= 
        \sum_{i=1}^3 \left|
        \sum_{j,k,l,m} \epsilon_{3jk}\,\epsilon_{ilm}
        \left( \sum_p (m_\nu)_{jp}\;(m_\nu)_{lp}^* - m_\pm \delta_{jl} \right)
        \left( \sum_q (m_\nu)_{kq}\;(m_\nu)_{mq}^* - m_\pm \delta_{km} \right)
        \right|^2.
    \end{aligned}
\end{equation}

The corresponding eigenvectors define the unitary matrix $U_\nu$, which together with the charged-lepton mixing matrix $U_l$ forms the Pontecorvo–Maki–Nakagawa–Sakata (PMNS) matrix 
\begin{equation}
 U_{\mathrm{PMNS}} =U_l^{\dagger}U_{\nu}=
\begin{cases}
\begin{pmatrix}
U_e^\dagger u_0 & U_e^\dagger u_- & U_e^\dagger u_+ \\[2pt]
U_{\mu}^\dagger u_0 & U_{\mu}^\dagger u_- & U_{\mu}^\dagger u_+ \\[2pt]
U_{\tau}^\dagger u_0 & U_{\tau}^\dagger u_- & U_{\tau}^\dagger u_+
\end{pmatrix} , & \text{for NO},\\[8pt]
\begin{pmatrix}
U_e^\dagger u_- & U_e^\dagger u_+ & U_e^\dagger u_0 \\[2pt]
U_{\mu}^\dagger u_- & U_{\mu}^\dagger u_+ & U_{\mu}^\dagger u_0 \\[2pt]
U_{\tau}^\dagger u_- & U_{\tau}^\dagger u_+ & U_{\tau}^\dagger u_0
\end{pmatrix} , & \text{for IO}.
\end{cases}
\end{equation}
In the standard parameterization, the PMNS matrix is written as
\begin{align}
    U_{PMNS} &=\left(\begin{array}{ccc}
c_{12} c_{13} & s_{12} c_{13} & s_{13} e^{-i \delta_{\mathrm{CP}}} \\
-s_{12} c_{23}-c_{12} s_{13} s_{23} e^{i \delta_{\mathrm{CP}}} & c_{12} c_{23}-s_{12} s_{13} s_{23} e^{i \delta_{\mathrm{CP}}} & c_{13} s_{23} \\
s_{12} s_{23}-c_{12} s_{13} c_{23} e^{i \delta_{\mathrm{CP}}} & -c_{12} s_{23}-s_{12} s_{13} c_{23} e^{i \delta_{\mathrm{CP}}} & c_{13} c_{23}
\end{array}\right)P,
\end{align}
where $c_{ij} = \cos \theta_{ij}$ and $s_{ij} = \sin \theta_{ij}$. The matrix $P = \operatorname{diag}(1, e^{i\sigma}, 1)$ contains the Majorana phase $\sigma$, which does not affect neutrino oscillations but contributes to neutrinoless double-beta decay.

From the components of the PMNS matrix, the lepton mixing angles can be determined as follows:
\begin{align}
\sin ^2\theta _{13}=\left | U_{e3} \right |^2=
\begin{cases}
\left | U_e^{\dagger }u_+ \right |^2   , & \text{for NO},\\[8pt]
\left | U_e^{\dagger }u_0 \right |^2 , & \text{for IO},
\end{cases}
\end{align}

\begin{equation}
\sin ^2\theta _{12}=\frac{\left | U_{e2} \right |^2 }{1-\left | U_{e3} \right |^2 } =
\begin{cases}
  \frac{|U_e^{\dagger }u_-|^2}{1-|U_e^{\dagger }u_+|^2}  , & \text{for NO},\\[8pt]
\frac{|U_e^{\dagger }u_+|^2}{1-|U_e^{\dagger }u_0|^2}  , & \text{for IO},
\end{cases}
\end{equation}

\begin{equation}
\sin ^2\theta _{23}=\frac{\left | U_{\mu 3} \right |^2 }{1-\left | U_{e3} \right |^2 } =
\begin{cases}
  \frac{|U_{\mu}^\dagger u_+|^2}{1-|U_e^\dagger u_+|^2} , & \text{for NO},\\[8pt]
\frac{|U_{\mu}^\dagger u_0|^2}{1-|U_e^\dagger u_0|^2}  , & \text{for IO}.
\end{cases}
\end{equation}

The Jarlskog invariant, which measures the magnitude of CP violation in the lepton sector, is given by
\begin{equation}
\begin{aligned}
J_{\mathrm{CP}} &= \mathrm{Im}\!\left[U_{\alpha i}U_{\beta j}U_{\alpha j}^*U_{\beta i}^*\right]
= c_{12}s_{12}c_{23}s_{23}c_{13}^2s_{13}\sin\delta \\
&=
\begin{cases}
  \mathrm{Im}\!\left[(U_e^\dagger u_0)(U_{\mu}^\dagger u_-)(U_e^\dagger u_-)^*(U_{\mu}^\dagger u_0)^*\right], & \text{for NO},\\[8pt]
  \mathrm{Im}\!\left[(U_e^\dagger u_-)(U_{\mu}^\dagger u_+)(U_e^\dagger u_+)^*(U_{\mu}^\dagger u_-)^*\right], & \text{for IO}.
\end{cases}
\end{aligned}
\end{equation}

The effective Majorana mass relevant for neutrinoless double-beta decay, $m_{\beta\beta}$, and the effective electron neutrino mass probed in beta-decay experiments, $m_{\beta}$, are given by
\begin{equation}
m_{\beta\beta}=\Big|\sum_i U_{ei}^2\,m_i\Big| =
\begin{cases}
  \Big|(U_e^\dagger u_-)^2\,m_-+(U_e^\dagger u_+)^2\,m_+\Big|  , & \text{for NO},\\[8pt]
\Big|(U_e^\dagger u_-)^2\,m_++(U_e^\dagger u_+)^2\,m_-\Big|  , & \text{for IO},
\end{cases}
\end{equation}
\begin{equation}
m_\beta=\sqrt{\sum_{i=1}^{3}|U_{ei}|^2\,m_i^2} =
\begin{cases}
  \sqrt{|U_e^\dagger u_-|^2\,m_-^2+|U_e^\dagger u_+|^2\,m_+^2}  , & \text{for NO},\\[8pt]
\sqrt{|U_e^\dagger u_-|^2\,m_+^2+|U_e^\dagger u_+|^2\,m_-^2} , & \text{for IO}.
\end{cases}
\end{equation}
These observables provide important probes of the neutrino mass scale and the Majorana nature of neutrinos. The predicted values of $m_{\beta\beta}$ and $m_{\beta}$ for the best-fit points are listed in table~\ref{tab:bestfit_combined}.

\subsection{ Scalar sector}
\label{subsec:sclrsct}

The scalar sector of the model consists of the SM Higgs doublet $H$, the singlet scalar $S$, and the $D_4$ doublet $\eta$. The most general scalar potential invariant under all symmetries is given by
\begin{equation}
    \begin{aligned}
V(H, S, \eta) = 
& -\mu_H^2 (H^\dagger H) + \lambda_H (H^\dagger H)^2 -\mu_S^2 S^\dagger S + \lambda_S (S^\dagger S)^2  + \lambda_{H,S} (H^\dagger H)(S^\dagger S) \\
&+ \lambda_{H,\eta} (H^\dagger H)\eta^2  - \mu_\eta^2 \eta^2 + \lambda_{S,\eta} (S^\dagger S)\eta^2 + \lambda_{\eta,1} (\eta^2)_{1_{++}}(\eta^2)_{1_{++}} 
   \\
   &+ \lambda_{\eta,2} (\eta^2)_{1_{+-}}(\eta^2)_{1_{+-}} 
   + \lambda_{\eta,3} (\eta^2)_{1_{-+}}(\eta^2)_{1_{-+}}.
\end{aligned}
\end{equation}
After expanding the scalars around their VEVs in the unitary gauge, as given in eq.~\eqref{eq:sclrvevexp}, the minimization conditions of the potential yield
\begin{equation}
    \begin{aligned}
   \mu_H^2=&\lambda _Hv^2+\frac{1}{2}\lambda _{H,S}u^2,\\
    \mu_S^2=&\lambda _Su^2+\frac{1}{2}\lambda _{H,S}v^2,\\
    \mu_{\eta}^2=&\frac{1}{2} (\lambda _{H,\eta}v^2+\lambda _{S,\eta}u^2),
\end{aligned}
\end{equation}
For the scalar potential to be bounded from below, the quartic couplings need to satisfy the following relations:
\begin{equation}
	\begin{aligned}
		&\lambda_H>0,\quad \lambda_S>0,\quad \lambda_{\eta,1}+\lambda_{\eta,2}>0,\quad \lambda_{H,S}>-2\sqrt{\lambda_H\lambda_S},\\[6pt]
		&4\lambda_H(\lambda_{\eta,1}+\lambda_{\eta,2})>\lambda_{H,\eta}^2,\quad
		4\lambda_S(\lambda_{\eta,1}+\lambda_{\eta,2})>\lambda_{S,\eta}^2,\quad \lambda_{\eta,2}+\lambda_{\eta,3}=0,	\end{aligned}
\end{equation}
where the last condition originates from $\partial V/\partial \eta_1=0$. The mass squared matrix for the neutral scalar fields in the basis $\left( \phi_H, \phi_S,\eta_2\right)$ is given by
\begin{align}
  M^2=  \begin{pmatrix}
 2\lambda_Hv^2 & \lambda _{H,S}uv   & \frac{\lambda _{H,\eta}vv_{\eta}}{\sqrt{2}} \\
 \lambda _{H,S}uv & 2\lambda_Su^2  & \frac{\lambda _{S,\eta}uv_{\eta}}{\sqrt{2}} \\
  \frac{\lambda _{H,\eta}vv_{\eta}}{\sqrt{2}} & \frac{\lambda _{S,\eta}uv_{\eta}}{\sqrt{2}}  &\left (4\lambda _{\eta,1}+4\lambda _{\eta,2}\right ) v_{\eta}^2
\end{pmatrix}. 
\label{eq:sclrmsmtrx}
\end{align}
The eigenvalues of this matrix correspond to the masses of the three neutral scalar bosons. Their expressions are analogous to those in eq.~\eqref{eq:CE}, with the corresponding coefficients $T_1^{\prime}$, $T_2^{\prime}$, $T_3^{\prime}$ are given by
\begin{equation}
	\begin{aligned}
		T_1^{\prime} &= 2\lambda_H v^2 + 2\lambda_S u^2 + (4\lambda_{\eta,1} + 4\lambda_{\eta,2})\, v_{\eta}^2,\\[6pt]
		T_2^{\prime} &= 4\lambda_H \lambda_S\, u^2 v^2
		+ 4\lambda_H (2\lambda_{\eta,1} + 2\lambda_{\eta,2})\, v_{\eta}^2 v^2
		+ 4\lambda_S (2\lambda_{\eta,1} + 2\lambda_{\eta,2})\, v_{\eta}^2 u^2 \\
		&\quad - 2\lambda_{S,\eta}^2\, u^2 v_{\eta}^2
		- \lambda_{H,S}^2\, u^2 v^2
		- 2\lambda_{H,\eta}^2\, v^2 v_{\eta}^2,\\[6pt]
		T_3^{\prime} &= 8\lambda_H \lambda_S (2\lambda_{\eta,1} + 2\lambda_{\eta,2})\, u^2 v^2 v_{\eta}^2
		- 4\lambda_H \lambda_{S,\eta}^2\, u^2 v^2 v_{\eta}^2 \\
		&\quad - 2(2\lambda_{\eta,1} + 2\lambda_{\eta,2}) \lambda_{H,S}^2\, u^2 v^2 v_{\eta}^2
		+ 4\lambda_{H,S} \lambda_{H,\eta} \lambda_{S,\eta}\, u^2 v^2 v_{\eta}^2 \\
		&\quad - 4\lambda_{H,\eta}^2 \lambda_S\, u^2 v^2 v_{\eta}^2.
	\end{aligned}
\end{equation}
The scalar mass squared matrix in eq.~\eqref{eq:sclrmsmtrx} exhibits a hierarchical structure dictated by the VEV hierarchy $v_\eta \gg v \sim u$. The 33 entry of the mass matrix  is larger than the off-diagonal entries in the third row and column, while the entries of the upper-left $2\times2$ block are the smallest. The calculation of the scalar mixing pattern applies procedures similar to those of the charged-lepton sector in eq.~\eqref{eq:clmm}, and the explicit diagonalization is not repeated here.

A comment on the scalar $\eta_1$ is in order. If the mass of $\eta_1$ were below the charged lepton masses, the charged leptons could decay into $\eta_1$. To forbid this decay channel, $m_{\eta_1}$ is raised to the GeV scale. This is achieved through the quartic term $\eta_1^2 \eta_2^2$ in the scalar potential, which induces a one-loop mass correction for $\eta_1$ via a $\eta_2$ loop. At the one-loop level, this quartic interaction does not generate sizable mass corrections for the other scalar fields $(\phi_H, \phi_S, \eta_2)$, leaving their masses essentially unaffected.

The scalar sector is subject to various constraints. Perturbativity requires all quartic couplings to remain below $4\pi$. Vacuum stability imposes positivity conditions on the quartic couplings. Furthermore, the masses of the new scalars must be consistent with constraints from collider searches for inert doublets and singlet scalars. While a detailed study of these constraints is beyond the scope of this work, the chosen parameter space ensures that the scalar sector does not conflict with existing experimental bounds.

\section{Phenomenology}
\label{sec:pheno}

In this section, the phenomenological implications of the model are examined in detail. The analysis is divided into three parts. First, the neutrino oscillation parameters are fitted to current experimental data, and the allowed parameter space is determined. Next, the viability of TeV-scale resonant leptogenesis is investigated, and the parameter regions that reproduce the observed baryon asymmetry are identified. Finally, the predictions for cLFV processes are computed and compared with the current experimental bounds.

\subsection{Neutrino masses and mixing}
\label{subsec:nuoscpara}

To fit the experimentally measured neutrino observables, a scan over the model parameter space is performed using the \texttt{flavorpy} package~\cite{Baur:2024flavorpy}. The best-fit values of the model parameters are determined by minimizing the chi-squared function
\begin{align}
    \chi^2=\sum_{i}\left ( \frac{p_i-q_i}{\sigma _i}  \right )^2,  
\end{align}
where $p_i$ denotes the theoretical prediction for a given observable, $q_i$ is the experimental central value, and $\sigma_i$ is the associated $1\sigma$ uncertainty. The sum runs over all neutrino observables: the mixing angles $\sin^2\theta_{12}$, $\sin^2\theta_{13}$, and $\sin^2\theta_{23}$, the charged-lepton mass ratios $m_e/m_\mu$ and $m_\mu/m_\tau$, the solar mass squared difference $\Delta m^2_{21}$, and the atmospheric mass squared difference $\Delta m^2_{3\ell}$. Here $\ell=1$ for NO and $\ell=2$ for IO.
\begin{table}[H]
	\centering
	\renewcommand{\arraystretch}{1.3} 
	\begin{tabular}{cccc}
		\toprule
		Parameter & Range & NO (bfp) & IO (bfp) \\
		\midrule
		$\alpha/\beta$   & $[-10^3,10^3]$ & $-111.53$ & $-141.55$ \\
		$k_1 v_\eta/(\beta\Lambda)$      & $[-10^3,10^3]$ & $-13.32$ & $-16.02$ \\
		$k_2 v_\eta/(\beta\Lambda)$      & $[-10^3,10^3]$ & $158.48$ & $-154.51$ \\    
		$y_1 v_\eta/(\Lambda Y_1)$        & $[-10,10]$  & $1.726$ & $0.216$ \\
		$M_1/M$          & $[-10,10]$  & $8.220-3.503i$ & $5.449+4.134i$ \\
		$M_2/M$          & $[-10,10]$  & $-0.304-0.101i$ & $-4.964+3.733i$ \\
		\bottomrule
	\end{tabular}
	\caption{Ranges of the dimensionless model parameters used in the numerical scan and their values obtained at the best-fit points (bfp). The first four parameters of the charged-lepton sector are real, whereas the last two are complex couplings in the neutrino sector. }
	\label{tab:parameters}
\end{table}

To perform the parameter scan, ranges are assigned to the dimensionless real parameters appearing in the mass matrices. The three real coupling ratios relevant to the charged-lepton sector are varied in the interval $[-10^3,\,10^3]$. The three coupling ratios associated with the neutrino sector, which include two complex parameters and one real parameter, are scanned over the range $[-10,\,10]$. These intervals are chosen to cover the natural values of the coupling constants while avoiding unnatural fine-tuning. The complete set of scanned parameters and their ranges is listed in table~\ref{tab:parameters}. 

With scanning of parameter space, the model successfully reproduces the observed neutrino mass hierarchies and mixing angles for both normal and inverted mass orderings. The best-fit points obtained from the scan are presented in table~\ref{tab:bestfit_combined}, together with the global-fit results for comparison. The minimum chi-squared values are $\chi^2_{\min}=0.01$ for NO and $\chi^2_{\min}=0.005$ for IO, indicating good agreement with the experimental data. For NO, the best-fit point favors the lower octant, while for IO it favors the upper octant. At the best-fit points, the Dirac CP phase is $\delta_{\mathrm{CP}} \simeq 8^\circ$ for NO and $\delta_{\mathrm{CP}} \simeq 33^\circ$ for IO. The sum of neutrino masses is predicted to be $\sum m_\nu \simeq 5.88\times 10^{-2}\,\mathrm{eV}$ for NO and $\sum m_\nu \simeq 9.89\times 10^{-2}\,\mathrm{eV}$ for IO, both consistent with the latest cosmological bounds.
\begin{table}[H]
    \centering
    
    \small
    \setlength{\tabcolsep}{4pt}
    \begin{tabular}{lcccc}
        \toprule
        Parameter & NO  & IO  & NO (bfp $\pm1\sigma$) & IO (bfp $\pm1\sigma$) \\
        \midrule
       $m_e/m_{\mu}$ & 0.00479 & 0.00479 & 0.00479 & 0.00479\\
        $m_{\mu}/m_{\tau}$ & 0.0565 & 0.0565 & 0.0565 & 0.0565 \\
        $\sin^2\theta_{23}$ & 0.470 & 0.550 & $0.470^{+0.017}_{-0.013}$ & $0.550^{+0.012}_{-0.015}$ \\
        $\sin^2\theta_{12}$ & 0.310 & 0.310 & $0.308^{+0.012}_{-0.011}$ & $0.308^{+0.012}_{-0.011}$ \\
        $\sin^2\theta_{13}$ & 0.02219 & 0.02231 & $0.02215^{+0.00056}_{-0.00058}$ & $0.02231^{+0.00056}_{-0.00056}$ \\
        $\Delta m_{21}^2$ /$10^{-5}$ eV$^2$ & 7.50 & 7.50 & $7.49^{+0.19}_{-0.19}$ & $7.49^{+0.19}_{-0.19}$ \\
        $\left|\Delta m_{3l}^2\right|$ /$10^{-3}$ eV$^2$ & 2.513 & 2.484 & $2.513^{+0.021}_{-0.019}$ & $2.484^{+0.020}_{-0.020}$ \\
        
        $\delta_{CP}$ ($^\circ$) & 8 & 33 & $212^{+26}_{-41}$ & $274^{+22}_{-25}$ \\
        $\sum m_{\nu}$/$10^{-2}$ eV & 5.878 & 9.890 & --- & --- \\
        $m_{\beta\beta}$/$10^{-3}$ eV & 3.6 & 48.2 & --- & ---\\
        $m_{\beta}$/$10^{-3}$ eV&8.9  & 49.8 & --- & ---\\
        $J_{CP}$ &0.00490 & 0.01840 & --- & ---\\
        \bottomrule
    \end{tabular}
     \caption{Best-fit values of the neutrino oscillation parameters predicted by the model at the minimum $\chi^2$ point, compared with the global-fit best-fit values and $1\sigma$ uncertainties from NuFit-6.0~\cite{Esteban:2024eli}. The minimum chi-squared values are $\chi^2_{\min}=0.01$ for NO and $\chi^2_{\min}=0.005$ for IO. Here $\ell=1$ denotes NO and $\ell=2$ denotes IO for the atmospheric mass squared difference $\Delta m^2_{3\ell}$.}
     \label{tab:bestfit_combined}
\end{table}

\begin{figure}[t] 
    \centering
    \includegraphics[width=1.0\textwidth]{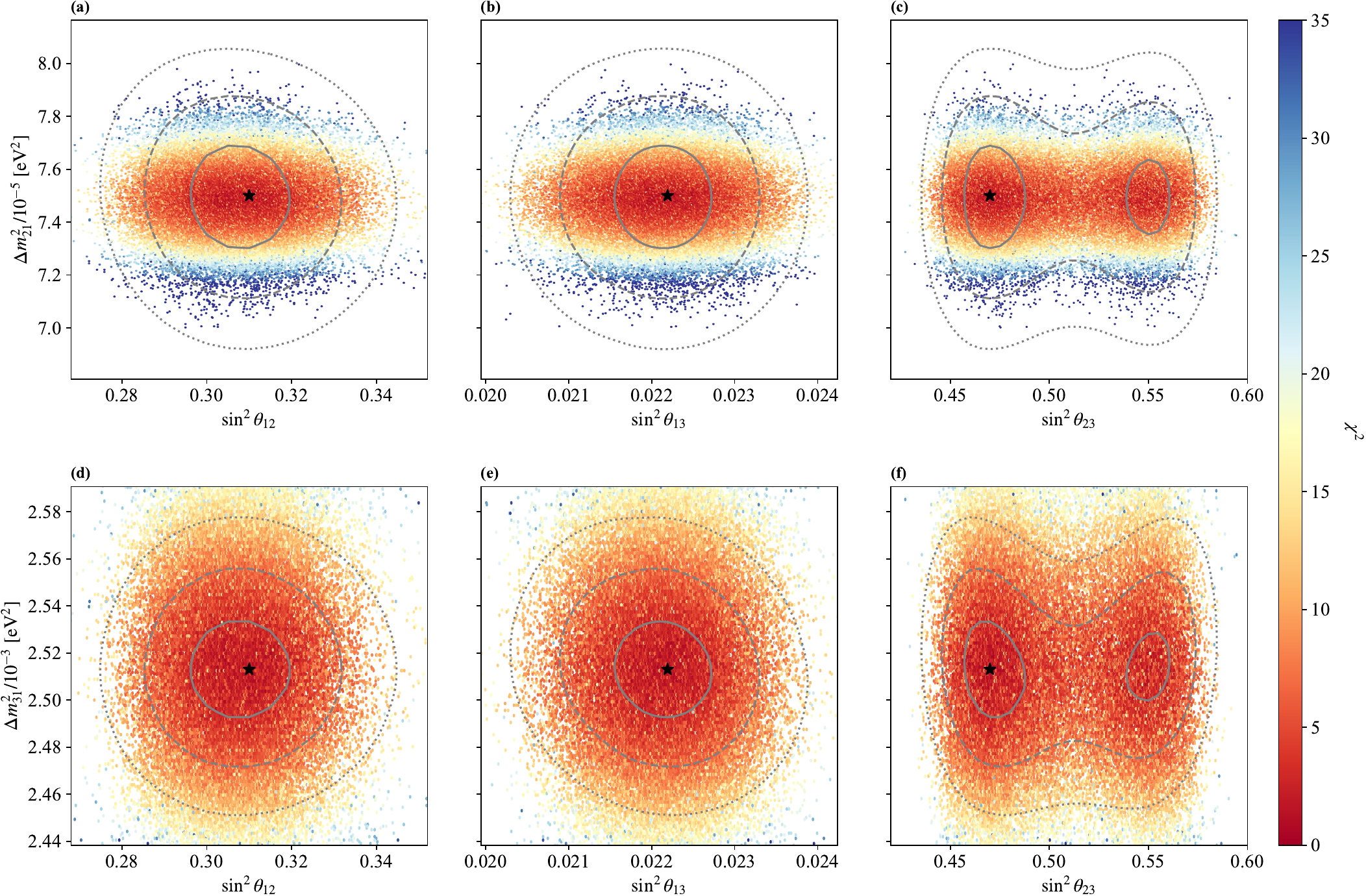}
    \caption{Two-dimensional $\chi^2$ distributions of the neutrino oscillation parameters for the NO. Panels (a)–(c) show the correlations of the solar mass squared difference $\Delta m^2_{21}$ with $\sin^2\theta_{12}$, $\sin^2\theta_{13}$, and $\sin^2\theta_{23}$, respectively. Panels (d)–(f) show the corresponding correlations for the atmospheric mass squared difference $\Delta m^2_{31}$. The color bar indicates the $\chi^2$ values. The gray contours denote the $1\sigma$, $2\sigma$, and $3\sigma$ experimental confidence regions. The black star marks the best-fit point of the model.}
    \label{fig:1}
\end{figure}

The two-dimensional $\chi^2$ distributions of the neutrino oscillation parameters for the NO scenario are displayed in figure~\ref{fig:1}. Figures~\ref{fig:1}(a)–\ref{fig:1}(c) show the correlations of the solar mass squared difference $\Delta m^2_{21}$ with the three mixing angles $\sin^2\theta_{12}$, $\sin^2\theta_{13}$, and $\sin^2\theta_{23}$, respectively. Figures~\ref{fig:1}(d)–\ref{fig:1}(f) show the corresponding correlations for the atmospheric mass squared difference $\Delta m^2_{31}$. The sampled points are color-coded by their $\chi^2$ values, ranging from 0 (red) to 35 (blue), as indicated by the color bar. The gray contours denote the experimental $1\sigma$, $2\sigma$, and $3\sigma$ confidence regions. The black star in each panel marks the best-fit point of the model. The predictions are in good agreement with the data, with the best-fit point lying within the $1\sigma$ allowed regions for all observables.

\begin{figure}[h] 
    \centering
    \includegraphics[width=1.0\textwidth]{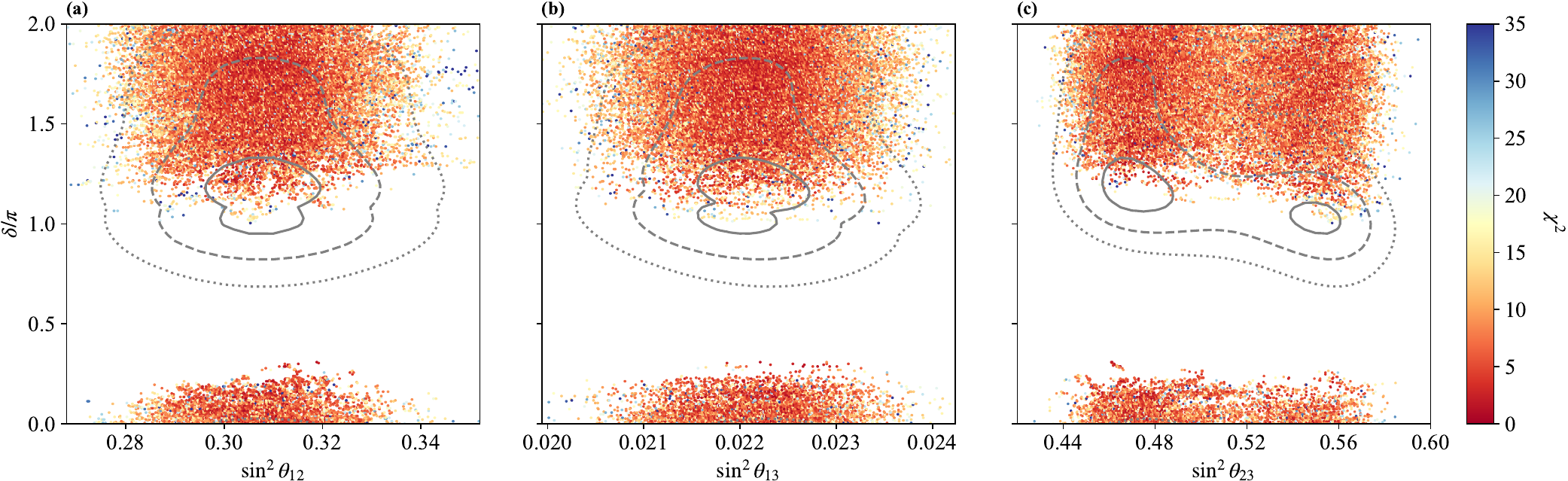}
    \caption{Two-dimensional $\chi^2$ distributions of the Dirac CP-violating phase $\delta/\pi$ versus the three neutrino mixing angles for the NO. Panel (a) shows $\delta/\pi$ versus $\sin^2\theta_{12}$, panel (b) versus $\sin^2\theta_{13}$, and panel (c) versus $\sin^2\theta_{23}$. The color coding and contours are the same as in figure~\ref{fig:1}.}
    \label{fig:2}
\end{figure}

Figure~\ref{fig:2} presents the two-dimensional $\chi^2$ distributions of the Dirac CP-violating phase $\delta/\pi$ versus the three neutrino mixing angles in the NO scenario. Figure \ref{fig:2}(a) shows a broad allowed region in the $\delta/\pi$ versus $\sin^2\theta_{12}$ plane, with low-$\chi^2$ points concentrated around $\sin^2\theta_{12}\simeq 0.28$–$0.34$ and $\delta/\pi\simeq 1.0$–$2.0$ or $0$–$0.2$. Figure \ref{fig:2}(b) displays a similar pattern for $\sin^2\theta_{13}$, where the favored region lies at $\sin^2\theta_{13}\simeq 0.021$–$0.023$. Figure \ref{fig:2}(c) reveals a broad distribution in the $\sin^2\theta_{23}$ plane, with favored points spanning $\sin^2\theta_{23}\simeq 0.44$–$0.56$. The bulk of the low-$\chi^2$ regions lie within the $3\sigma$ experimental contours, indicating good agreement with the current constraints.

\begin{figure}[htbp] 
    \centering
    \includegraphics[width=1.0\textwidth]{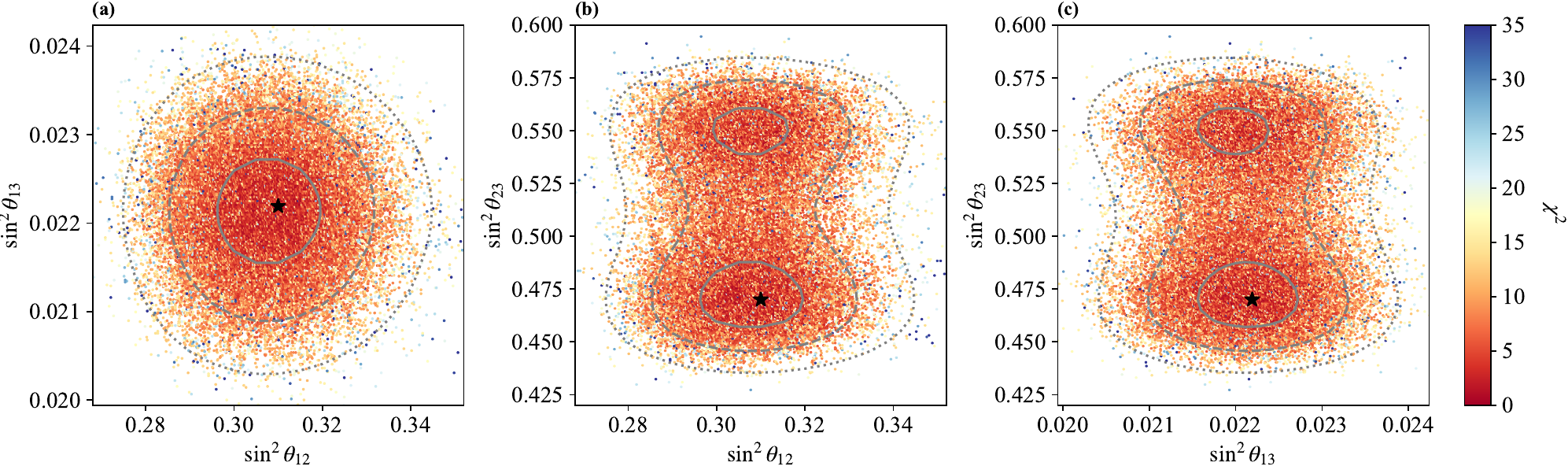}
    \caption{Two-dimensional $\chi^2$ distributions among the three neutrino mixing angles for the NO. Panel (a) shows the correlation between $\sin^2\theta_{12}$ and $\sin^2\theta_{13}$, panel (b) between $\sin^2\theta_{12}$ and $\sin^2\theta_{23}$, and panel (c) between $\sin^2\theta_{13}$ and $\sin^2\theta_{23}$. The color coding and contours are the same as in figure~\ref{fig:1}. The black star marks the best-fit point, which lies within the $1\sigma$ allowed region for all three correlations. }
    \label{fig:3}
\end{figure}

Figure~\ref{fig:3} shows the two-dimensional $\chi^2$ distributions among the three neutrino mixing angles in the NO scenario. Figure~\ref{fig:3}(a) displays the correlation between $\sin^2\theta_{12}$ and $\sin^2\theta_{13}$, figure~\ref{fig:3}(b) between $\sin^2\theta_{12}$ and $\sin^2\theta_{23}$, and figure~\ref{fig:3}(c) between $\sin^2\theta_{13}$ and $\sin^2\theta_{23}$. The model predictions are well compatible with the current experimental constraints. The best-fit point lies within the $1\sigma$ allowed region for all three correlations. In the NO scenario, the best-fit point favors the lower octant, although the overall distribution of points spans both octants.

\begin{figure}[t] 
    \centering
    \includegraphics[width=1.0\textwidth]{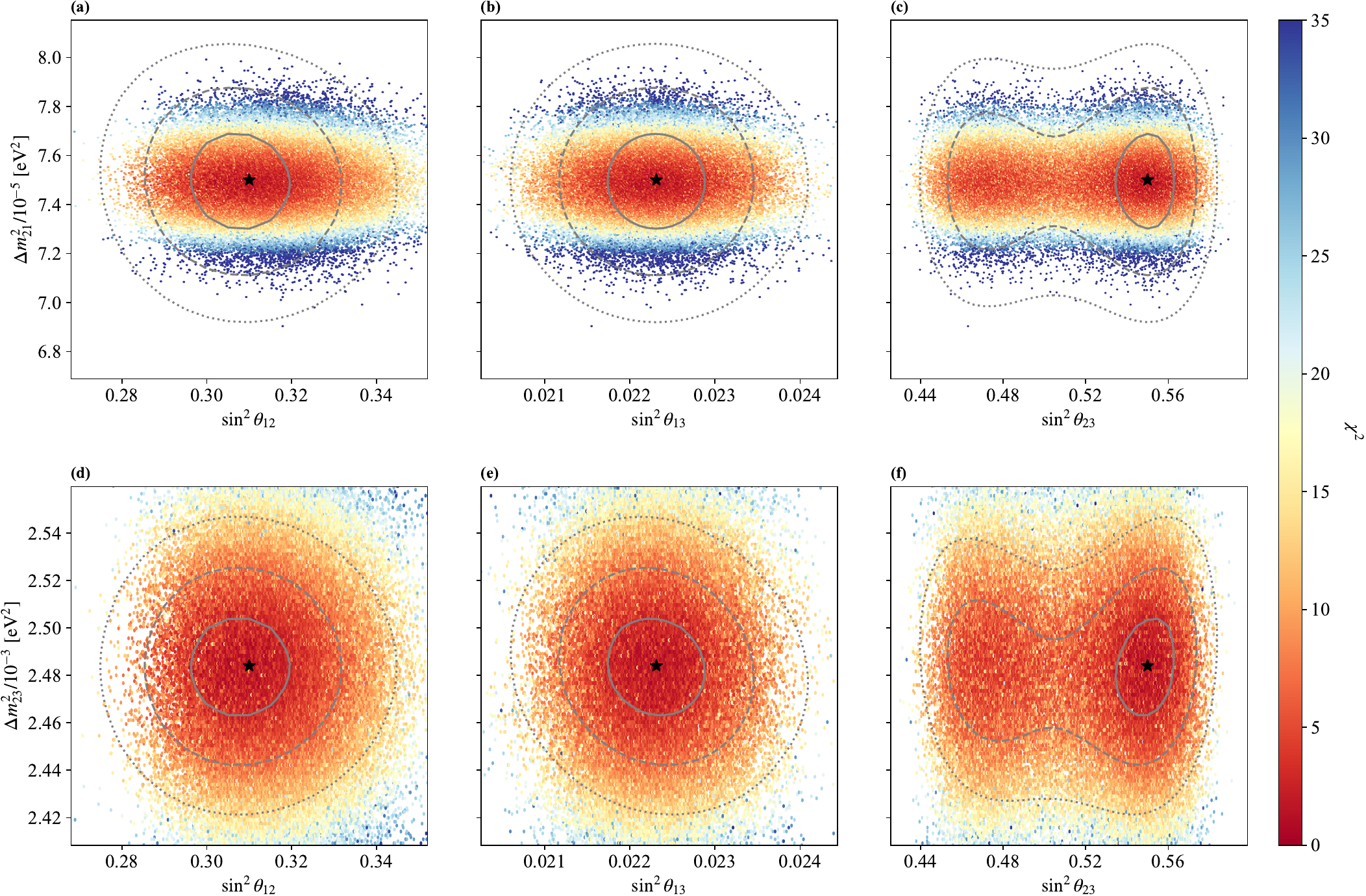}
    \caption{Two-dimensional $\chi^2$ distributions of the neutrino oscillation parameters for the IO. Panels (a)–(c) show the correlations of the solar mass squared difference $\Delta m^2_{21}$ with $\sin^2\theta_{12}$, $\sin^2\theta_{13}$, and $\sin^2\theta_{23}$, respectively. Panels (d)–(f) show the corresponding correlations for the atmospheric mass squared difference $\Delta m^2_{32}$. The color coding and contours are the same as in figure~\ref{fig:1}. The black star marks the best-fit point, lying within the $3\sigma$ allowed regions for all observables. The model predictions are in good agreement with the data.}
    \label{fig:16}
\end{figure}

Figure~\ref{fig:16} presents the two-dimensional $\chi^2$ distributions of the neutrino oscillation parameters for the IO scenario. Figures \ref{fig:16}(a)–\ref{fig:16}(c) show the correlations of the solar mass squared difference $\Delta m^2_{21}$ with $\sin^2\theta_{12}$, $\sin^2\theta_{13}$, and $\sin^2\theta_{23}$, respectively. Figures \ref{fig:16}(d)–\ref{fig:16}(f) show the corresponding correlations for the atmospheric mass squared difference $\Delta m^2_{32}$. The model predictions are again in good agreement with the data, with the best-fit point lying well within the $3\sigma$ allowed regions for all observables. Unlike the NO case, the best-fit point in the IO scenario favors the upper octant, although the data points in panels involving $\sin^2\theta_{23}$ show a spread across both octants.

\begin{figure}[t] 
    \centering
    \includegraphics[width=1.0\textwidth]{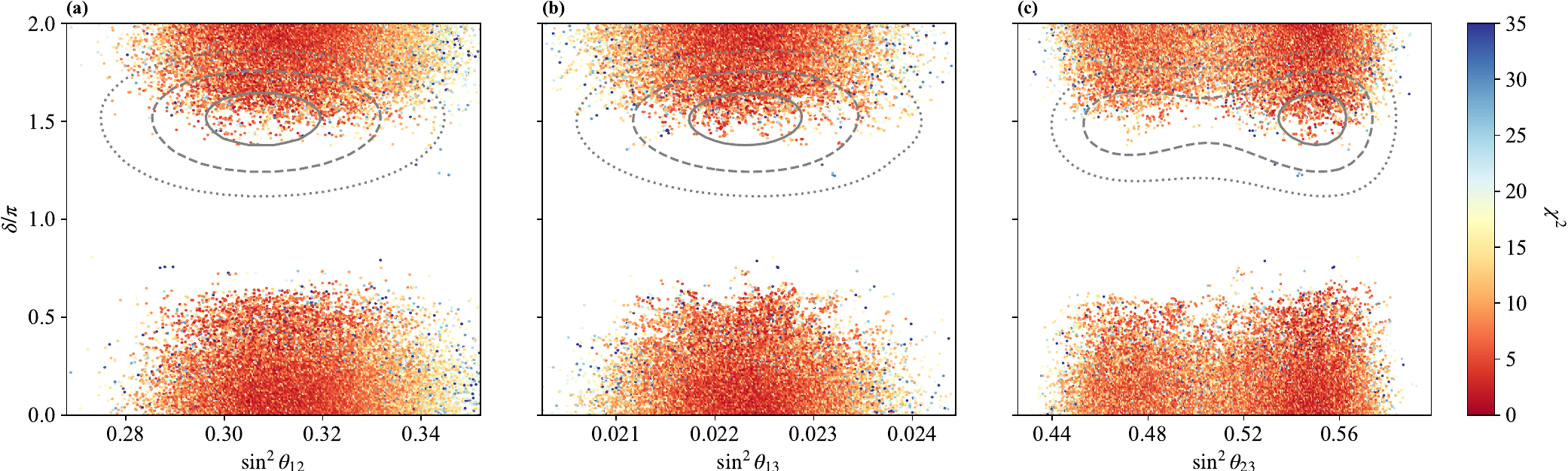}
    \caption{The $\chi^2$ distributions of the Dirac CP-violating phase $\delta$ versus the three neutrino mixing angles for the IO. Panel (a) shows $\delta/\pi$ versus $\sin^2\theta_{12}$, panel (b) versus $\sin^2\theta_{13}$, and panel (c) versus $\sin^2\theta_{23}$. The color coding and contours are the same as in figure~\ref{fig:1}. The black star is the position of the best-fit point.}
    \label{fig:5}
\end{figure}

Figure~\ref{fig:5} presents the  $\chi^2$ distributions of the Dirac CP-violating phase $\delta/\pi$ versus the three neutrino mixing angles for the IO scenario. Figure~\ref{fig:5}(a) shows $\delta/\pi$ versus $\sin^2\theta_{12}$, figure~\ref{fig:5}(b) versus $\sin^2\theta_{13}$, and figure~\ref{fig:5}(c) versus $\sin^2\theta_{23}$. The low-$\chi^2$ regions are roughly concentrated around $\delta/\pi \sim 0$–$0.6$ and $1.5$–$2.0$. Similar to the NO case, the constraints on the CP phase in the IO scenario are also weaker, as reflected by the broader distribution of allowed points.

\begin{figure}[htbp] 
    \centering
    \includegraphics[width=1.0\textwidth]{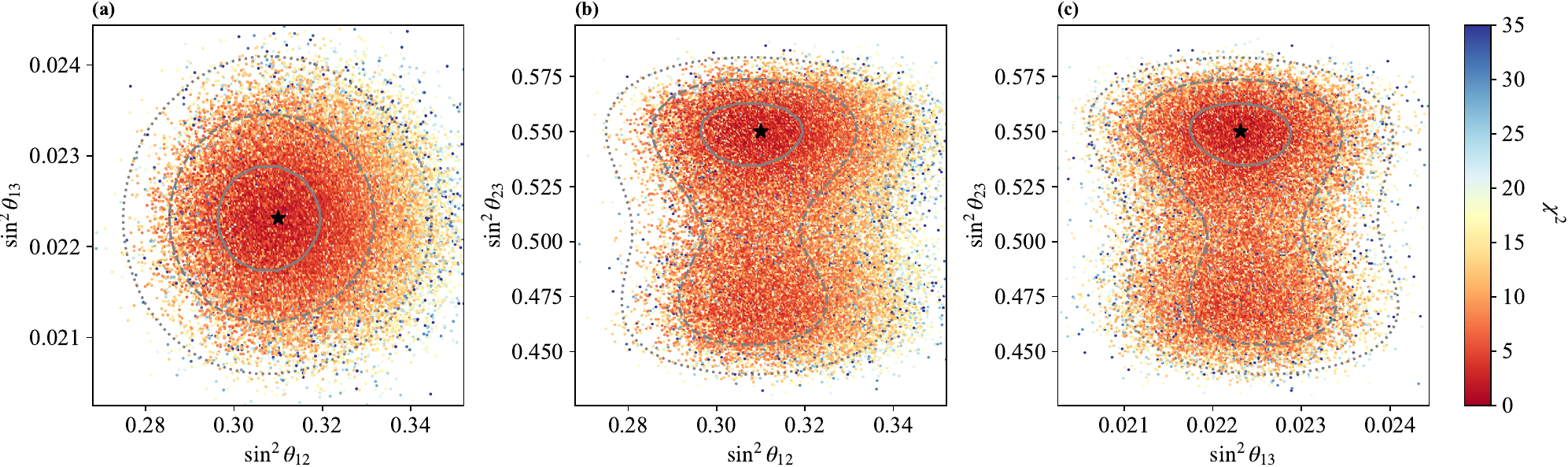}
    \caption{The $\chi^2$ distributions among the three neutrino mixing angles for the IO. Panel (a) shows the correlation between $\sin^2\theta_{12}$ and $\sin^2\theta_{13}$, panel (b) between $\sin^2\theta_{12}$ and $\sin^2\theta_{23}$, and panel (c) between $\sin^2\theta_{13}$ and $\sin^2\theta_{23}$. The color coding and contours are the same as in figure~\ref{fig:1}. The black star is the best-fit point. The favored region in the $\sin^2\theta_{12}$–$\sin^2\theta_{13}$ plane is relatively concentrated, while the panels involving $\sin^2\theta_{23}$ show a broader distribution.}
    \label{fig:6}
\end{figure}

Figure~\ref{fig:6} displays $\chi^2$ distributions among the three neutrino mixing angles for the IO scenario. Figure~\ref{fig:6}(a) shows the correlation between $\sin^2\theta_{12}$ and $\sin^2\theta_{13}$, figure~\ref{fig:6}(b) between $\sin^2\theta_{12}$ and $\sin^2\theta_{23}$, and figure~\ref{fig:6}(c) between $\sin^2\theta_{13}$ and $\sin^2\theta_{23}$. The best-fit points for all three correlations lie inside the $1\sigma$ allowed regions. The favored region in the $\sin^2\theta_{12}$–$\sin^2\theta_{13}$ plane is relatively concentrated, while the panels involving $\sin^2\theta_{23}$ show a broader distribution. In particular, the best-fit points in figures~\ref{fig:6}(b) and ~\ref{fig:6}(c) prefer the upper octant for $\theta_{23}$, consistent with the conclusion drawn from figure~\ref{fig:16}(c).

\begin{figure}[t] 
	\centering
	\includegraphics[width=1.0\textwidth]{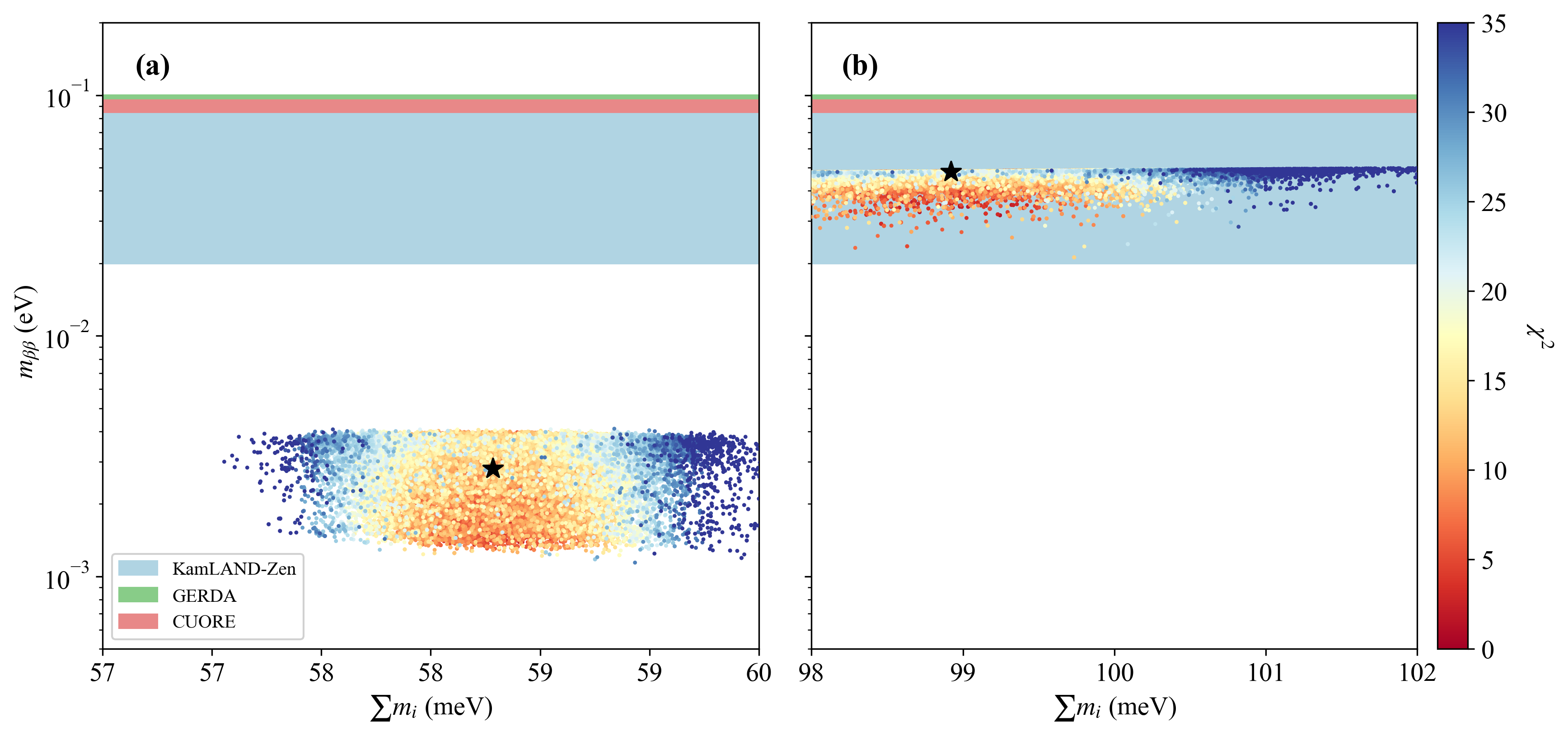}
	\caption{Correlation between the sum of neutrino masses $\sum m_i$ and the effective Majorana mass $m_{\beta\beta}$ for (a) the NO and (b) the IO. The color coding and contours are the same as in figure~\ref{fig:1}, and the black star is the best-fit point.}
	\label{fig:7c}
\end{figure}

The correlations between the effective Majorana mass $m_{\beta\beta}$ and the sum of neutrino masses $\sum m_i$ are displayed in figure~\ref{fig:7c}, for both neutrino mass orderings. In the NO scenario, shown in figure~\ref{fig:7c}(a), the model predicts $m_{\beta\beta}$ to lie in the range of $1$--$4$\,meV, which is well below the sensitivity reach of current experiments. In contrast, for the IO case, shown in figure~\ref{fig:7c}(b), the predicted values fall within the range of $20$--$50$\,meV. This region lies within the sensitivity reach of the current KamLAND‑Zen experiment~\cite{KamLAND-Zen:2024eml}. This places tight constraints on the parameter space, as indicated by the upper horizontal shaded band. The green horizontal band and the red horizontal band represent the current exclusion limits from the GERDA~\cite{Agostini:2020jxc} and CUORE~\cite{Adams:2020yqy} experiments, respectively. All sampled points satisfy $\sum m_i < 0.12$\,eV~\cite{Planck:2018vyg,GAMBITCosmologyWorkgroup:2020rmf}, consistent with the latest cosmological bounds~\cite{GAMBITCosmologyWorkgroup:2020rmf}.  The IO scenario can be further scrutinized by the next‑generation neutrinoless double‑beta decay experiments, such as nEXO~\cite{nEXO:2021ujk}, LEGEND‑1000~\cite{LEGEND:2021bnm}, and CUPID~\cite{CUPID:2019imh}, which are expected to either confirm or rule out the model predictions in the future.

From the analyses above, it can be seen that the model provides a good fit to the current neutrino oscillation data for both normal and inverted mass orderings. The best-fit points lie within the $1\sigma$ experimental regions for all observables. The NO scenario favors the lower octant for $\theta_{23}$, while the IO scenario favors the upper octant. At the best fit point, the Dirac CP phases are $\delta_{\mathrm{CP}} \simeq 8^\circ$ for NO and $\delta_{\mathrm{CP}} \simeq 33^\circ$ for IO. The  neutrinoless double-beta decay observable $m_{\beta\beta}$ lies below the current experimental limits, with the IO scenario falling in the reach of next-generation experiments. These results demonstrate that the model is a viable and testable framework for understanding the neutrino mass and mixing pattern.

\subsection{TeV-scale Leptogenesis}
\label{subsec:lowsaclleptognss}

In this subsection, the generation of the baryon asymmetry through leptogenesis is investigated within the ISS(2,2) framework. As discussed in section~\ref{subsec:yukfermass}, the ISS mechanism naturally accommodates nearly degenerate heavy neutrino pairs. These pairs can resonantly enhance the CP asymmetry in their decays, enabling a sizable lepton asymmetry to be generated at the TeV scale. This asymmetry is subsequently converted into the baryon asymmetry via electroweak sphaleron processes. The leptogenesis analysis presented below focuses on the parameter regions that are consistent with the neutrino oscillation best-fit points obtained in the previous subsection, and identifies the conditions under which the observed baryon asymmetry of the Universe can be reproduced.

In the basis $(N,\chi)$, the $4\times 4$ mass matrix for heavy neutrinos takes the block form
\begin{align}
    M_H=\begin{pmatrix}
 0 &M_R \\
  M_R^T&\mu
\end{pmatrix} ,
\label{eq:MH}
\end{align}
which is diagonalized by a unitary transformation $V$ as
 \begin{align}
    V^TM_HV=\mathrm{diag}\left (M_{N_1},M_{N_2},M_{N_3},M_{N_4}\right ).
    \label{eq:diagMH}
\end{align}
The resulting mass eigenstates form two almost pseudo-Dirac pairs, $(N_1,N_2)$ and $(N_3,N_4)$, with nearly degenerate masses. The small mass splitting within each pair is generated by the parameter $\mu$, whose smallness is important for resonantly enhancing the CP asymmetry. In the numerical analysis that follows, only the lighter pair $(N_1,N_2)$ is considered, as its contribution to the lepton asymmetry dominates over that of the heavier pair, as the contribution of the latter is strongly suppressed by washout effects~\cite{Chakraborty:2021azg}.

For the decay of a heavy neutrino $N_i$ ($i=1,2$) into a lepton doublet $l_\alpha$ and the Higgs doublet $H$, the CP asymmetry $\epsilon_{i\alpha}$ is given by~\cite{Covi:1996fm,Blanchet:2009kk}
\begin{align}
\epsilon_{i \alpha}=\frac{1}{8 \pi} \sum_{j \neq i}\left\{\frac{\operatorname{Im}\left[h_{i \alpha}^{*} h_{j \alpha}\left(\sum_{\gamma} h_{i \gamma}^{*} h_{j \gamma}\right)\right]}{\sum_{\beta}\left|h_{i \beta}\right|^{2}} f^{v}_{i j}+\frac{\operatorname{Im}\left[h_{i \alpha}^{*} h_{j \alpha}\left(\sum_{\gamma} h_{i \gamma} h_{j \gamma}^{*}\right)\right]}{\sum_{\beta}\left|h_{i \beta}\right|^{2}} f^{c}_{i j}\right\},
\label{eq:cpasym}
\end{align}
where $f^v_{ij}$ denotes the loop factor from self-energy and vertex diagrams that violate lepton number, and $f^c_{ij}$ that from self-energy diagrams that conserve it.  Expressions for these functions are derived in Refs.~\cite{Covi:1996fm,Covi:1996wh,Blanchet:2010kw}. In the nearly degenerate limit $M_{N_1}\simeq M_{N_2}$, the loop functions satisfy  $f_{21}^v\simeq -f_{12}^v\simeq f_{21}^c\simeq -f_{12}^c\equiv f_{ij}$, where $f_{ij}$ is defined as~\cite{Covi:1996wh,Buchmuller:1997yu,Blanchet:2009kk}
\begin{align}
	f_{ij}=\frac{\left ( M^2_{N_i}- M^2_{N_j}\right )M_{N_i}M_{N_j} }{\left ( M^2_{N_i}- M^2_{N_j} \right )^2+\left ( M_{N_i}\Gamma_i-M_{N_j}\Gamma_j\right ) ^2 }.
	\label{eq:resfunc}
\end{align}
The function $f_{ij}$ is resonantly enhanced when the mass splitting within the pair becomes comparable to the decay width. The enhancement is largest for nearly degenerate pseudo-Dirac pairs, where the small mass splitting generated by $\mu$ allows the CP asymmetry to reach values far above those obtained in the conventional seesaw. When lepton number is exactly conserved, $f_{ij}$ diverges~\cite{BhupalDev:2014pfm}, whereas the CP-violating prefactor vanishes~\cite{Blanchet:2009kk}. The decay width of $N_i$ entering the denominator of eq.~\eqref{eq:resfunc} is
\begin{align}
	\Gamma_i=\frac{\left ( hh^\dagger  \right )_{ii}M_{N_i} }{8\pi }.
	\label{eq:decwidth}
\end{align}
The Yukawa couplings $h_{I\alpha}$ , with $I=1,2,\dots 4$, in the mass basis are related to those in the flavor basis through the diagonalization matrix $V$ as follows
\begin{equation}
    \begin{aligned}
    h_{I\alpha }=y_{\alpha 1}V_{1I}^*+y_{\alpha 2}V_{2I}^*.
\end{aligned}
\label{eq:dagnol}
\end{equation}
The evolution of the number densities is governed by the Boltzmann equations~\cite{Buchmuller:1997yu,Agashe:2018cuf}
\begin{equation}
    \begin{aligned}
    \frac{\mathrm{d} N_{i}}{\mathrm{d} z}&=-D_i\left ( N_{i}-N^{eq}_{i} \right ),\qquad i=1,2\\
    \frac{\mathrm{d}N_{B-L}^{\alpha }}{\mathrm{d}z}&=-\sum_{i=1}^{2}\epsilon _{i\alpha } D_i\left ( N_{i}-N^{eq}_{i} \right )-\sum_{i=1}^{2} P_{\alpha i }W_i N_{B-L}^{\alpha }\delta_{i} ^2,
    \end{aligned}
    \label{eq:system} 
\end{equation}
where $N_{i}$, and $N_{B-L}$ denote the comoving number densities,  $N_{i}^{\rm eq}$ is the equilibrium number density, $z=M_{N_1}/T$, and $\delta_i=\left | \Delta M \right |/\Gamma_i$ with $\Delta M$ being the mass splitting within each pair of pseudo-Dirac sterile neutrinos. The equilibrium densities are $N_{i}^{\rm eq}(z_i)=(3/8)z_i^2K_2(z_i)$, with $i=1, 2$. $H$ is the Hubble expansion rate, and $W$ is the washout term. The term $D_i$ accounts for decays and inverse decays~\cite{Buchmuller:2004nz}
\begin{equation}
	D_i \equiv \frac{\Gamma_{i}}{H z}=K_i z\frac{K_1(z)}{K_2(z)},
\end{equation}
where $K_1(z)$ and $K_2(z)$ are the modified Bessel functions. The decay parameter $K_i$ and the Hubble expansion rate at $T=M_{N_i}$ are given by
\begin{align}
	K_i=\frac{\Gamma_i}{H(M_{N_i})},\qquad H(M_{N_i})=\sqrt{\frac{4\pi ^3g_*}{45} } \frac{M_{N_i}^2}{M_\mathrm{Pl}},
\end{align}
where $g_{*}$ is the effective degrees of freedom and the Planck mass is $M_{\mathrm{Pl}} = 1.22 \times 10^{19} \ \mathrm{GeV}$. The flavor projection operator appearing in the last term of eq.~\eqref{eq:system} is defined as
\begin{align}
	P_{\alpha i}=\frac{\left| h_{\alpha i} \right|^{2} }{\left (h^{\dagger }h\right )_{ii}},\qquad \sum_{\alpha}P_{\alpha i}=1.
	\label{eq:flavproj}
\end{align}

The solution for $N_{B-L}$ from eq.~\eqref{eq:system} can be written as the sum of two terms:
\begin{align}
    N_{B-L}(z)=\sum_{i,\alpha }N_{B-L}^{i,\alpha }e^{-\int_{z_i}^{z}P_{\alpha i}W_i(z')\delta _i^2dz' }-\frac{3}{4}\sum_{i,\alpha }\epsilon _{i\alpha }\kappa _{i\alpha }, \label{eq:NBL_solution} 
\end{align}
where $N_{B-L}^{i,\alpha}$ is the $\alpha$ flavor initial $B-L$ asymmetry, which is set to zero in our numerical analysis. The first term therefore vanishes, and the second term describes the $B-L$ production from $N_i$ decays. It is expressed in terms of the efficiency factor $\kappa_{i\alpha}$, which does not depend on the CP asymmetry $\epsilon_{i\alpha}$. The efficiency factor for each neutrino, which quantifies the fraction of the generated asymmetry that survives washout, is given by
\begin{equation}
    \begin{aligned}
\kappa_{i\alpha} (z) 
& =-\frac{4}{3} \int_{z_{\mathrm{i}}}^{z} d z^{\prime}  \frac{d N_{i}}{d z^{\prime}} e^{-\int_{z^{\prime}}^{z} d z^{\prime \prime} P_{\alpha i}W_i\left(z^{\prime \prime}\right)\delta_i^2}.
\end{aligned}
\label{eq:approxint}
\end{equation}

The baryon asymmetry is obtained from the $B-L$ asymmetry through sphaleron conversion and dilution. The relation between $N_{B-L}$ and the ratio of the baryon number density to the photon number density reads $\eta_B\simeq 1.28\times10^{-2}N_{B-L}$~\cite{Buchmuller:1997yu}. Also, the baryon-to-photon ratio and baryon-to-entropy ratio satisfy $\eta_B\simeq 7.04\,Y_B$~\cite{Davidson:2008bu}. It follows from eq.~\eqref{eq:NBL_solution} that the baryon number density to the entropy density gives
\begin{equation}
	Y_B\simeq-1.36\times 10^{-3}\sum_{i\alpha}\epsilon_{i\alpha}\kappa_{i\alpha},
\end{equation}
which is consistent with the existing result $Y_B\simeq -1.33\times 10^{-3}\sum_{i\alpha}\epsilon_{i\alpha}\kappa_{i\alpha}$~\cite{Shao:2026bgc,Shao:2025xav}.

To determine the baryon asymmetry, it is necessary to examine the efficiency factor in eq.~\eqref{eq:approxint} in detail. It clearly depends on the washout
\begin{equation}
	W_i = \frac{1}{4}K^{\mathrm{eff}}_i z_i^3 K_1(z_i).
\end{equation}
The effective washout parameter is~\cite{Blanchet:2009kk,Shao:2026bgc,Shao:2025xav}
\begin{align}
	K^{\mathrm{eff}}_i=
	\frac{K_i \delta_i^2}{1+\sqrt{a_i}\delta_i +\delta_i^2 } \simeq K_i\delta_i^2,
\end{align}
where $a_i=\left ( \Gamma_i /M \right )^2$, and the last equality holds for $\delta_i \ll 1$, which is satisfied in our setup. The expression for the final efficiency factor corresponding to all ranges of the effective washout parameter $K_\mathrm{eff}$ for thermally produced heavy neutrinos is given below~\cite{Buchmuller:2004nz}:
\begin{equation}
    \kappa_{i\alpha}(z) =
\begin{cases}
\displaystyle \frac{2}{K_{i\alpha}^{\mathrm{eff}}\,z}\,
\Bigl[\,1 - \exp\!\Bigl(-\frac{K_{i\alpha}^{\mathrm{eff}}}{2}\,\Phi(z)\Bigr)\Bigr], & z < z_B, \\[12pt]
\displaystyle \frac{2}{K_{i\alpha}^{\mathrm{eff}}\,z_B}\,
\Bigl[\,1 - \exp\!\Bigl(-\frac{K_{i\alpha}^{\mathrm{eff}}}{2}\,I(z)\Bigr)\Bigr], & z \ge z_B,
\end{cases}
\label{eq:pweta}   
\end{equation}
where
\begin{equation}
	\begin{aligned}
		\Phi(z) &= \int_{z_0}^{z} x^3 K_1(x)\,dx\simeq\frac{3\pi z^3}{\left [(9\pi  )^{0.7}+(2z^3)^{0.7}\right ]^{1.43}} ,\\
		I(z) &= \int_{z_0}^z x^2 K_1(x)\,dx = 2 - z^2 K_0(z) - 2z K_1(z),
	\end{aligned}
\end{equation}
where $z_B$ is well approximated by~\cite{Buchmuller:1997yu}
\begin{align}
	z_{B} \approx 1+\frac{1}{2} \ln \left[1+\frac{\pi (K_i^{\mathrm{eff}})^2}{2^{10}}\left(5\ln{\frac{5}{4}+\ln\left ( \pi (K_i^{\mathrm{eff}})^2 \right )  } \right)^{5}\right] .
\end{align}

Having established the theoretical framework for the leptogenesis, we now turn to its numerical analysis. In our analysis, we neglect contributions from scattering processes, thermal corrections, and other subleading effects. The light neutrino oscillation parameters are fixed to the model-predicted best-fit values listed in table~\ref{tab:bestfit_combined}. We scan the parameters over physically reasonable ranges, with heavy neutrinos at the TeV scale and lighter singlet states at the keV scale, such that the model reproduces the Planck observed baryon asymmetry $Y_B = 8.72 \times 10^{-11}$~\cite{Planck:2018vyg}.

To avoid introducing an excessive number of free parameters, we factor out a common scaling factor from $M_D$, $\mu$, and $M$ in our parameter scan~\footnote{There is an abuse of notation for $\mu$. Although it is defined as a matrix in eq.~\eqref{eq:masmatinv}, hereafter we mean its overall scale $\mu_0$. }. Within the ISS framework, this universal scaling factor can be expressed as
\begin{equation}
	n = \frac{Y_1^2 v^2 \mu}{2M^2}.
\end{equation}
The value of $n$ is fixed by requiring that the model reproduce the observed light neutrino masses, and is obtained from the parameter scan. For fixed $n$, the parameters  $Y_1$, $\mu$, and $M$ entering the mass matrices may be rescaled in ways that leave the neutrino oscillation observables unchanged. We adopt the scaling $Y_1 \to r Y_1$ and $M \to r M$, under which the combination $Y^2_1/M^2$ appearing in $n$ remains invariant. At $r = 1$, the reference values are set to $M=10~\mathrm{TeV}$ and $\mu=100~\mathrm{keV}$, which together with $n$ fix the Yukawa coupling $Y_1$. The scale parameter $r$ is then scanned over a wide range, and the baryon asymmetry yield $Y_B$ is computed for both NO and IO. Under this scaling transformation, the Yukawa couplings in eq.~\eqref{eq:dagnol} and the decay widths in eq.~\eqref{eq:decwidth} transform as $h \to rh$ and $\Gamma \to r^3\Gamma$, respectively. The CP asymmetry in eq.~\eqref{eq:cpasym} can then be written in terms of the rescaled quantities as
\begin{equation}
\begin{split}
\epsilon_{i\alpha}(r)
&\simeq -\frac{r^2}{8\pi}
\sum_{j\ne i=1}^{2}
\frac{\mathrm{Im}\left[
h_{i\alpha}^{*}h_{j\alpha}
\left(hh^\dagger+h^\dagger h\right)_{ij}
\right]}
{\sum_{\beta}|h_{i\beta}|^2}\,
f_{ij}(r),\\
f_{ij}(r)
&= \frac{(M_{N_i}^2-M_{N_j}^2)M_{N_i}M_{N_j}}
{(M_{N_i}^2-M_{N_j}^2)^2
+r^4(M_{N_i}\Gamma_i-M_{N_j}\Gamma_j)^2},
\end{split}
\label{eq:rescaledcpasym}
\end{equation}
where $M_{N_i}$, $h$, $\Gamma_i$, etc., are the corresponding physical quantities evaluated at $r=1$. This expression describes the enhancement of the lepton asymmetry as $r$ varies.

\begin{figure}[t]
	\includegraphics[width=\textwidth]{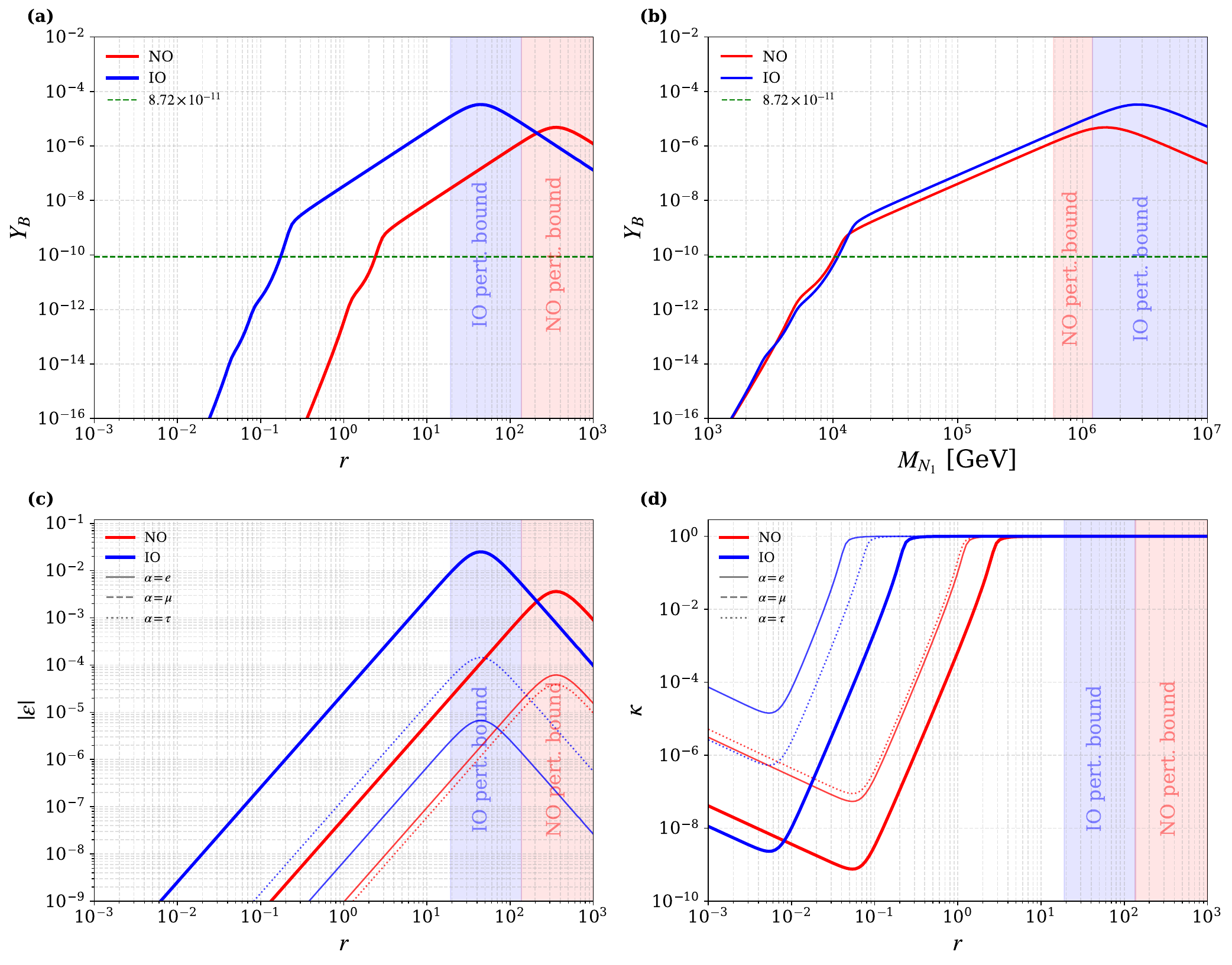}
	\caption{Baryon asymmetry and its underlying quantities as functions of the scaling parameter $r$ for NO (red curve) and IO (blue curve). (a) Baryon asymmetry $Y_B$ as a function of $r$. The horizontal dashed line is the value observed by Planck, $Y_B = 8.72 \times 10^{-11}$~\cite{Planck:2018vyg}. The shaded regions denote parameter space excluded by perturbativity of the Yukawa couplings, with light blue for IO and light red for NO. (b) $Y_B$ as a function of the lightest pseudo-Dirac neutrino mass $M_{N_1}$. (c) CP-violating parameter $\epsilon$ as a function of $r$. (d) Efficiency factor $\kappa$ of the lightest pseudo-Dirac pair as a function of $r$.    }
	\label{fig:scheme1_summary}
\end{figure}

The rescaling behavior encoded in eq.~\eqref{eq:rescaledcpasym} provides the theoretical guideline for the phenomenological analysis that follows. It shows that the single parameter $r$ controls the competition between the enhancement of the CP asymmetry and the suppression of the washout, so the observed baryon asymmetry can naturally be achieved over a broad range of $r$. We therefore scan $r$  over a wide range and compute $Y_B$ for both NO and IO. The results are summarized in figure~\ref{fig:scheme1_summary}.

Figure~\ref{fig:scheme1_summary}(a) shows the baryon asymmetry $Y_B$ as a function of $r$, while figure~\ref{fig:scheme1_summary}(b) displays $Y_B$ as a function of the lightest pseudo-Dirac neutrino mass $M_{N_1}$, which scales linearly with $r$. In both figures, the horizontal dashed line indicates the observed baryon asymmetry by the Planck collaboration.  Successful leptogenesis requires pseudo-Dirac masses around 10 TeV for both NO and IO.

The shape of $Y_B(r)$ follows from the competing $r$ dependence of the washout and the CP asymmetry. As $r$ increases, the washout parameter $K_i$ grows linearly with $r$, whereas the mass-degeneracy parameter $\delta_i=|\Delta M|/\Gamma_i\sim\mu/\Gamma_i$ falls as $r^{-3}$, since $\Gamma_i\propto r^3$. The effective washout parameter $K_i^{\rm eff}\sim K_i\delta^2_i$ therefore decreases as $r^{-5}$, and the efficiency factor rises as $r^{5}$,  as shown in figure~\ref{fig:scheme1_summary}(d). This rapid suppression of washout is the primary reason for the initial steep rise of the baryon asymmetry.

Figure~\ref{fig:scheme1_summary}(c) shows the CP-violating parameter $\epsilon$ as a function of $r$. As expected from eq.~\eqref{eq:rescaledcpasym}, $\epsilon$ exhibits a resonant peak when the mass splitting becomes comparable to the decay width, i.e., when the denominator of eq.~\eqref{eq:rescaledcpasym} is minimized. For small $r$, the mass splitting term dominates and $\epsilon$ grows as $r^2$, whereas for large $r$, the decay-width term dominates and $\epsilon$ falls as $r^{-2}$. The position of this peak depends on the neutrino mass ordering and the model parameters. The interplay between this resonant enhancement of $\epsilon$ and the $r^5$ growth of the efficiency factor determines the final shape of $Y_B(r)$ shown in figure~\ref{fig:scheme1_summary}(a). Flavor effects are included in these results. The $\mu$-flavor contribution dominates over the $e$ and $\tau$ contributions, so that the total CP asymmetry and the total efficiency factor closely follow their $\mu$-flavor counterparts. This feature reflects the structure of the $M_D$ mass matrix.

In the IO scenario, the predicted baryon asymmetry rises rapidly in the region $r<0.1$, crosses the observed value within $r\in(0.1,0.2)$, and reaches its peak $Y_B\approx10^{-4}$ at $r\simeq40$–$50$. Beyond the peak, $Y_B$ decreases gradually with increasing $r$. The red curve corresponding to NO is shifted toward larger values of $r$. It intersects the Planck line near $r\simeq 2.5$, peaks at $Y_B\approx10^{-5}$ in the interval $r\simeq300$–$400$, and then declines continuously. The shaded regions on the right-hand side of all panels denote the excluded parameter space where the Yukawa couplings violate perturbativity, with the light-blue region corresponding to the IO exclusion and the light-red region to the NO exclusion. These viable mass ranges lie in the multi-TeV regime, far below the characteristic scale of the conventional type-I seesaw mechanism. This suggests that the heavy pseudo-Dirac neutrinos predicted by our model are promising targets for collider searches, and it demonstrates the phenomenological viability of low-scale leptogenesis within the ISS framework.
\begin{figure}[htbp] 
    \centering
    \includegraphics[width=1.0\textwidth]{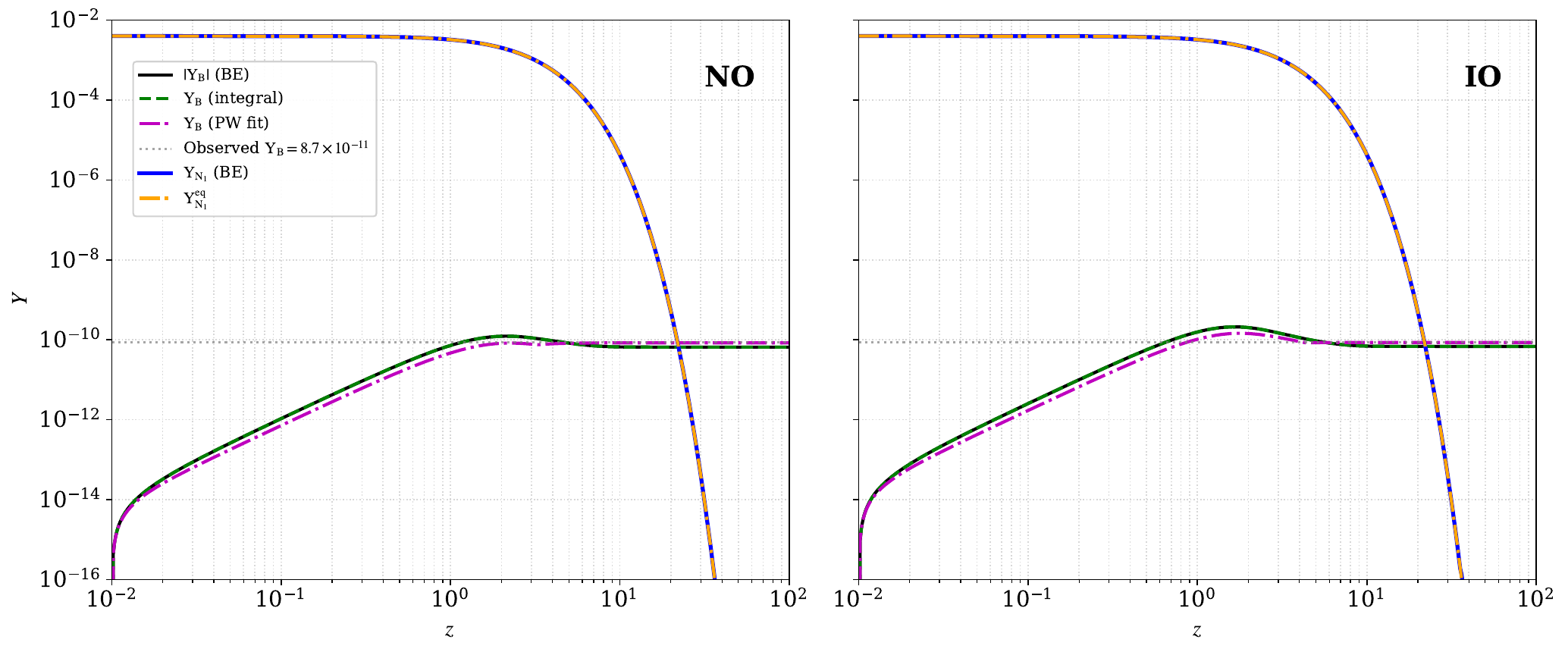}
    \caption{Evolution of the baryon asymmetry yield $|Y_B|$ and the lightest heavy-neutrino abundance $Y_{N_1}$ as functions of $z$, for representative parameter points in the NO (left panel, $r=r_{\rm NO}$) and IO (right panel, $r=r_{\rm IO}$) scenarios. The black solid curve is the full numerical solution of the Boltzmann equations, eq.~\eqref{eq:system}. The green dashed and magenta dash-dotted curves are obtained from the integral approximation, eq.~\eqref{eq:approxint}, and the piecewise approximation, eq.~\eqref{eq:pweta}, respectively. The blue solid and orange dash-dotted curves correspond to the numerical and equilibrium abundances $Y_{N_1}$ and $Y_{N_1}^{\rm eq}$. The gray dotted horizontal line marks the observed value $Y_B=8.72\times10^{-11}$~\cite{Planck:2018vyg}. }
    \label{fig:9}
\end{figure}

Figure~\ref{fig:9} illustrates the evolution of the heavy-neutrino number density and the generated lepton asymmetry as functions of the dimensionless parameter $z$, for representative choices of the scale parameter $r$ in the scaling, for both  NO and IO of the neutrino mass spectrum. The vertical axis denotes the comoving number density. The left panel corresponds to NO with $r_{\rm NO}=2.4$, where the lightest heavy neutrino capable of generating a sizable lepton asymmetry has a mass around $10\ \mathrm{TeV}$. The right panel corresponds to IO with $r_{\rm IO}=0.17$, where the lightest heavy neutrino mass is approximately $10\ \mathrm{TeV}$. In both panels, the black solid line shows the baryon asymmetry yield $|Y_B|$ obtained from numerical solutions of the full Boltzmann equations, eq.~\eqref{eq:system}. The green dashed and magenta dash-dotted lines correspond to $|Y_B|$ derived from the integral analytic formula, eq.~\eqref{eq:approxint}, and the piecewise approximation, eq.~\eqref{eq:pweta}, respectively. The gray dotted line indicates the observed value $Y_B = 8.72\times10^{-11}$ from the Planck satellite~\cite{Planck:2018vyg}. The blue solid line represents the numerical abundance $Y_{N_1}$ of the lightest heavy neutrino, while the orange dash-dotted line shows its equilibrium value $Y_{N_1}^{\mathrm{eq}}$.  Since the Yukawa interactions associated with the heavy neutrino are strong, its abundance closely tracks its thermal equilibrium value throughout its evolution. Therefore, we adopt the approximate relation $Y_{N_1} \approx Y_{N_1}^{\rm eq}$.   

The lepton asymmetry begins its equilibrium evolution as $z$ approaches zero. In contrast, as the universe cools, the system departs from thermal equilibrium, and $|Y_B|$ reaches its maximum around $z \simeq 1$. Subsequently, washout processes become dominant, and $|Y_B|$ gradually decreases, eventually settling at the level of $10^{-11}$. For $z \gg 1$, the cosmic temperature falls below the production threshold of heavy neutrinos, and the baryon asymmetry freezes out. The evolution shown in the right panel is qualitatively similar to the NO case, both of which lie in the weak‑washout regime.

As can be seen in the figure~\ref{fig:9}, the evolution of $Y_B$ is governed by the competition between two classes of physical processes: CP-violating decays act as the source of the asymmetry, while washout processes such as inverse decays and scattering tend to erase it. The analytic approximations reproduce the numerical results with high accuracy. The green dashed line and the magenta dash-dotted line, corresponding to the integral approximation and the piecewise analytic fit, respectively, track the full numerical solution over the entire evolution range, confirming the reliability of both treatments.

The scaling analysis above followed a single direction, $Y_1 \to r Y_1$ with $M \to r M$ at fixed $n$. A broader scan is also possible. Defining  $k = Y_1 v / M$,  the combination $n$ can be written as $n=k^2\mu/2$. For fixed $n$, each value of $\mu$ therefore fixes $k$. Scanning $\mu$ over the range $[1, 10^4]\,\text{keV}$ covers the allowed values of $k$. At each such value of $k$, points are sampled randomly in the two-dimensional parameter space $(Y_1 v, M)$. The points are subject to the hierarchy $M_R \gg M_D > \mu$, and those with Yukawa couplings above the perturbativity bound $\sqrt{4\pi}$ are discarded.

\begin{figure}[t] 
	\centering
	\includegraphics[width=1.0\textwidth]{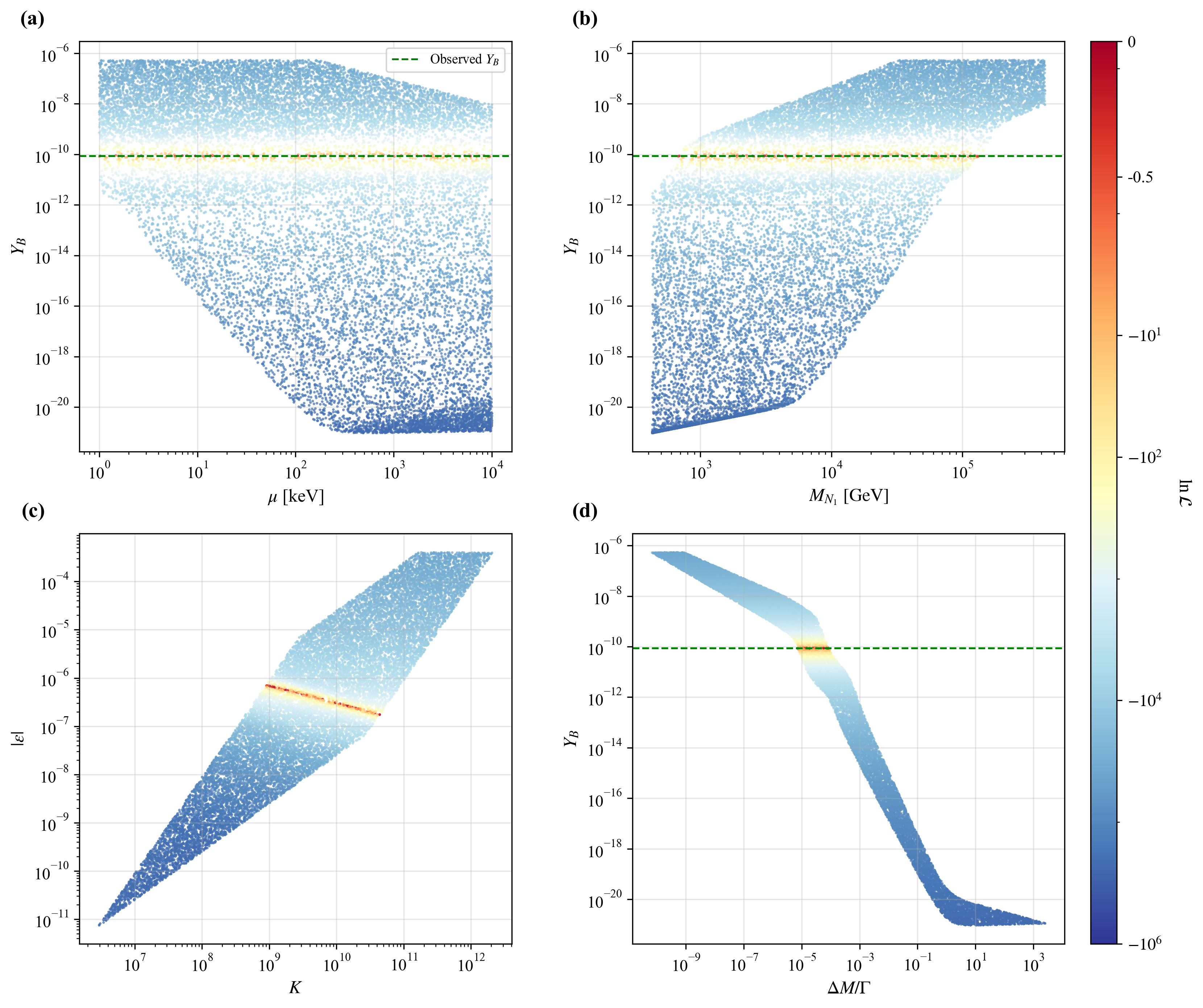}
	\caption{Viable parameter space for successful leptogenesis in the NO scenario. Each panel shows the predicted baryon asymmetry $\eta_B$ as a function of a different model parameter: (a) the sterile neutrino scale parameter $\mu$; (b) the heavy neutrino mass $M_{N_1}$; (c) CP asymmetry $\epsilon$ versus washout parameter $K_i$; (d) the ratio of the mass splitting to the decay width, $\Delta M/\Gamma$. The horizontal dashed line in each panel marks the observed value $\eta_B^{\rm obs}$.}
	\label{fig:15a}
\end{figure}

To map out the parameter space that reproduces the observed baryon asymmetry, a Bayesian scan is performed using the \texttt{dynesty} nested sampling package~\cite{Speagle:2019ivv}. The log-likelihood function used in the scan is
\begin{align}
	\ln\mathcal{L} = -\frac{1}{2}\left(\frac{\eta_B^{i} - \eta_B^{\mathrm{obs}}}{\Delta\eta_B^{\mathrm{obs}}}\right)^{2},
\end{align}
where $\eta_B^{i}$ is the baryon asymmetry predicted by the model at a given parameter point, $\eta_B^{\mathrm{obs}} = 6.12\times10^{-10}$ is the observed baryon asymmetry of the Universe, and $\Delta\eta_B^{\mathrm{obs}} = 0.04\times10^{-10}$ is its $1\sigma$ uncertainty~\cite{Planck:2018vyg}. The results are shown in figure~\ref{fig:15a}.

As shown in figure~\ref{fig:15a}(a), the singlet neutrino mass parameter $\mu$ can reproduce the cosmologically observed baryon asymmetry over a broad range, $1~\mathrm{keV}$--$10~\mathrm{MeV}$. Figure~\ref{fig:15a}(b) shows that viable leptogenesis can be realized for pseudo-Dirac pair masses in the range $1$--$100~\mathrm{TeV}$. Figure~\ref{fig:15a}(c) gives the washout parameter $K_i$ in the range $10^9$--$10^{11}$, while the CP asymmetry $\epsilon$ lies roughly in the range $10^{-7}$--$10^{-6}$. Figure~\ref{fig:15a}(d) further shows that a baryon asymmetry, consistent with the observed value, can be produced when the ratio of the mass splitting to the decay width, $\Delta M/\Gamma$, lies between $10^{-5}$ and $10^{-4}$. Taken together, the four panels show that the final asymmetry is determined jointly by the lepton-number-violating source and washout effects.

\subsection{Charged lepton flavor violation}
\label{subsec:clfv}

Beyond the neutrino masses, mixing, and leptogenesis, the model also predicts cLFV processes. These processes provide complementary constraints on the model, since they are induced at the one-loop level by the same heavy neutrino and $W$-boson exchanges that control leptogenesis, and they depend on the heavy–light mixing in a way that is largely independent of the low-energy neutrino observables.

The branching ratios for the cLFV processes $\mu\to e\gamma$, $\tau\to e\gamma$, and $\tau\to\mu\gamma$ in the ISS(2,2) model are given by~\cite{Ilakovac:1994kj,Hisano:1995cp,Deppisch:2004fa}
\begin{align}
	\mathrm{BR}\left(\ell_{\alpha} \rightarrow \ell_{\beta} \gamma\right)
	=\frac{\alpha_{\mathrm{em}}^{3} \sin^{2}\theta_{W}}{2^8\pi^{2}m_{W}^{4}}
	\frac{m_{\ell_{\alpha}}^{5}}{\Gamma_{\ell_{\alpha}}}
	\left|\sum_{I=1}^{4} U_{\alpha I} U_{\beta I}^{*}\,
	F\!\left(\frac{M_{N_I}^{2}}{m_{W}^{2}}\right)\right|^{2},
\end{align}
where $m_{\ell_\alpha}$ and $\Gamma_{\ell_\alpha}$ are the mass and total decay width of the initial-state charged lepton, $M_{N_I}$ are the masses of the heavy neutrinos, $\alpha_{\mathrm{em}}$ is the fine-structure constant, $m_W$ is the $W$-boson mass, and $\theta_W$ is the weak mixing angle. The decay widths of the initial leptons are taken as $\Gamma_\mu = 2.996\times10^{-19}$ GeV and $\Gamma_\tau = 2.267\times10^{-12}$ GeV~\cite{ParticleDataGroup:2026mpi}. The heavy–light mixing is defined as~\cite{Vien:2025mvm}
\begin{align}
	U_{\alpha I}\approx\frac{v}{\sqrt{2}}\frac{h_{\alpha I}}{M_{N_I}},
	\label{eq:U}
\end{align}
with $v=246\ \mathrm{GeV}$, and $h_{\alpha I}$ denoting the Yukawa couplings in the mass basis. The loop function $F(x)$, with $x=M_{N_I}^2/m_W^2$, is given by~\cite{Ilakovac:1994kj,Hisano:1995cp,Deppisch:2004fa}
\begin{align}
	F(x)=\frac{1}{6\left(x-1\right)^{4}}\left\{18\ln x+(x-1)\left[x\left(33+\left(4x-45\right)x\right)-10\right]
	\right\}.
\end{align}

\begin{figure}[htbp] 
    \centering
    \includegraphics[width=1.0\textwidth]{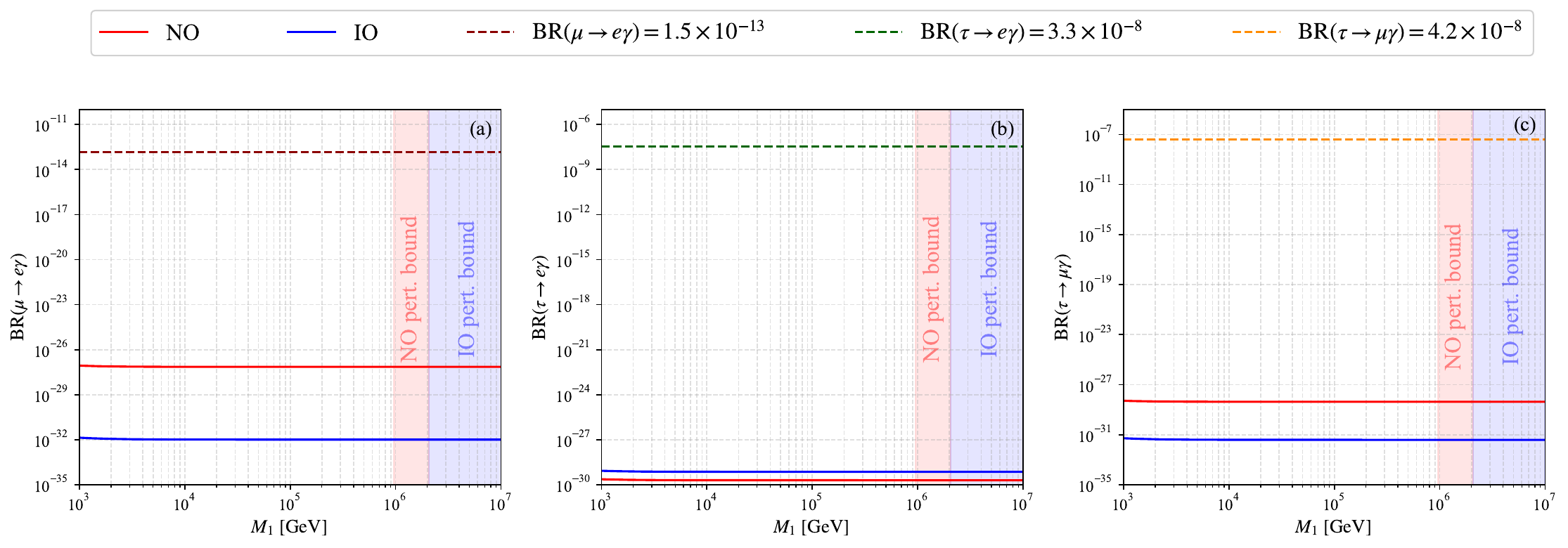}
    \caption{Branching ratios of the cLFV processes (a) $\mu\to e\gamma$, (b) $\tau\to e\gamma$, and (c) $\tau\to\mu\gamma$ as functions of the mass $M_{N_1}$ of the lightest pseudo-Dirac pair, under the scaling factor $r$. The red and blue solid curves correspond to NO and IO, respectively. The horizontal dashed lines denote the current experimental upper limits on each channel~\cite{Arkani-Hamed:2026wwy}, while the vertical shaded bands indicate the perturbativity bounds on the Yukawa coupling for the two mass orderings.}
    \label{fig:10}
\end{figure}

Figure~\ref{fig:10} displays the theoretical predictions for the branching ratios of the cLFV decay channels as a function of the scaling factor $r$, plotted against the mass $M_{N_1}$ of the lightest pseudo-Dirac pair. The experimental bounds are $\mathrm{BR}(\mu\to e\gamma) < 1.5\times10^{-13}$ at 90\% C.L. by MEG II~\cite{MEGII:2023ltw}, $\mathrm{BR}(\tau\to e\gamma) < 3.3\times10^{-8}$ at 90\% C.L. by BaBar~\cite{BaBar:2009hkt}, and  $\mathrm{BR}(\tau\to\mu\gamma) < 4.2\times10^{-8}$ at 90\% C.L. by Belle~\cite{Belle:2021ysv}. The red and blue solid curves correspond to the results of NO and IO, respectively. The horizontal dashed lines indicate the current experimental upper limits on each channel~\cite{Arkani-Hamed:2026wwy}, while the vertical shaded bands mark the perturbativity bounds on the Yukawa coupling for the two mass orderings.

In all the panels of figure~\ref{fig:10}, the active–sterile neutrino mixing satisfies the scaling relation $U\sim Y_1 v/M$, as shown in eqs.~\eqref{eq:U} and~\eqref{eq:dagnol}. Since both $Y_1$ and $M$ are rescaled by the same factor $r$, the mixing angle remains approximately unchanged. Moreover, the loop function $F(M_{N_1}^2/m_W^2)$ is evaluated entirely in the heavy-mass regime $M_{N_1}\gg m_W$, where it depends only logarithmically on the mass parameter. Consequently, over the entire scanned parameter space the cLFV branching ratios exhibit a very weak dependence on $M_{N_1}$, and the curves are nearly horizontal.

For $\mu\to e\gamma$ and $\tau\to\mu\gamma$, the NO predictions are several orders of magnitude larger than those for IO. This stems from the fact that, during the rotation of the couplings from the flavor basis to the mass basis in eq.~\eqref{eq:dagnol}, the relative sign between the contributions of the lighter and heavier pseudo-Dirac pairs flips in the IO case, leading to a cancellation in the amplitude. In contrast, the $\tau\to e\gamma$ prediction is independent of the phases of $h_{I\alpha}$, and NO yields a slightly larger branching ratio than IO for the parameter space considered here.

From figure~\ref{fig:10}, it is evident that the branching ratios predicted by the model lie many orders of magnitude below the current experimental sensitivities. Specifically, the branching ratio of $\mu\to e\gamma$ falls within the range $10^{-32}$–$10^{-27}$, while the branching ratio of $\tau\to e\gamma$ lies in the range $10^{-30}$–$10^{-29}$ and that of $\tau\to\mu\gamma$ lies in $10^{-32}$–$10^{-27}$.  
\begin{figure}[ht] 
    \centering
    \includegraphics[width=1.0\textwidth]{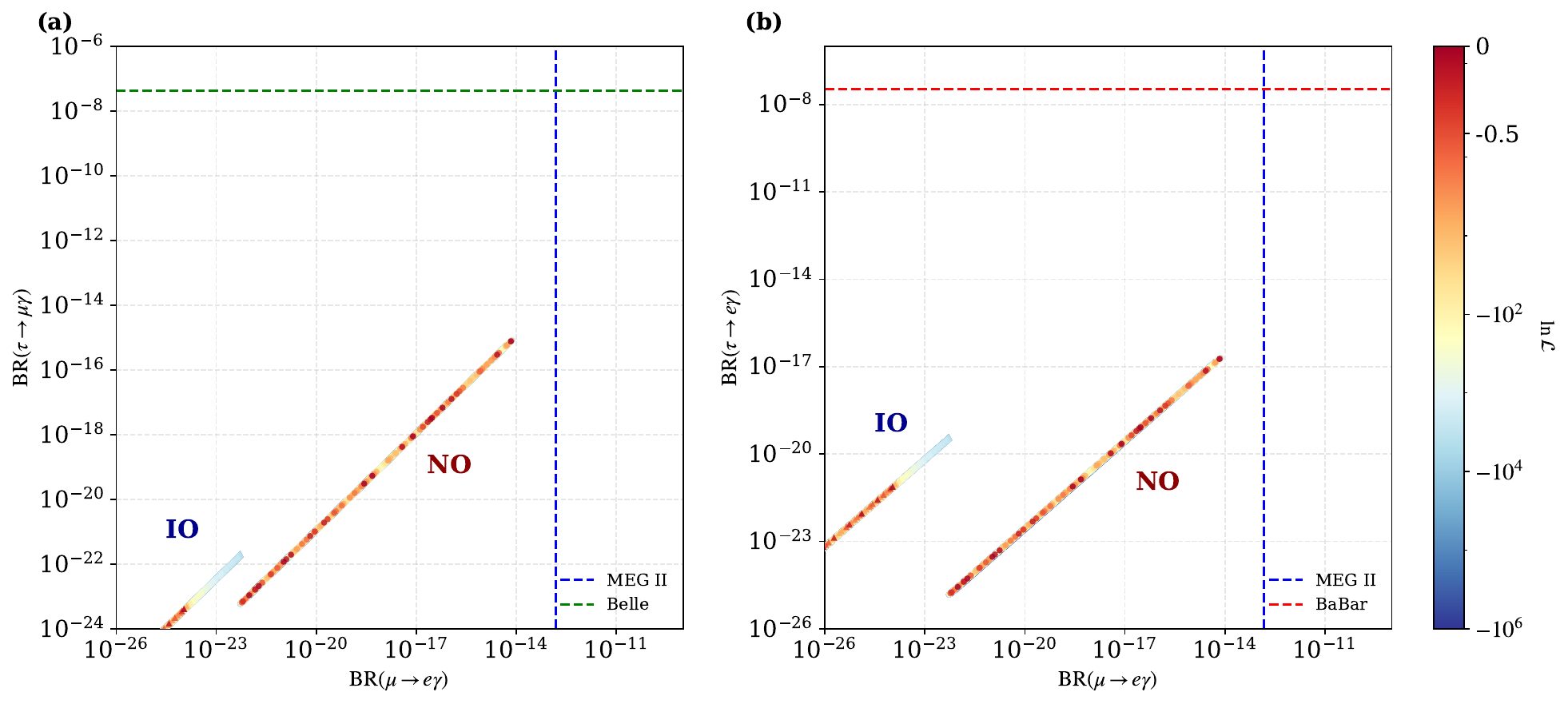}
	\caption{
	Scatter plots between the branching ratios of the cLFV processes. Panel (a) for $\mathrm{BR}(\tau\to\mu\gamma)$ versus $\mathrm{BR}(\mu\to e\gamma)$, and (b) for $\mathrm{BR}(\tau\to e\gamma)$ versus $\mathrm{BR}(\mu\to e\gamma)$. The color scale encodes the value of the log-likelihood $\ln \mathcal{L}$. The vertical dashed line is the MEG~II upper limit on $\mathrm{BR}(\mu\to e\gamma)$, while the horizontal dashed lines are the BaBar limit (red) on the $\tau\to e \gamma$ and Belle limit (green) on the $\tau\to \mu \gamma$. 
	}
    \label{fig:15b}
\end{figure}

Furthermore, the cLFV branching ratios for both NO and IO cases, in the parameter space consistent with the observed baryon asymmetry, are illustrated in figure~\ref{fig:15b}. The numerical results show that, within the parameter space that successfully reproduces the observed baryon asymmetry, the branching ratios of the three decays $\tau\to\mu\gamma$, $\tau\to e\gamma$, and $\mu\to e\gamma$ all lie below the current experimental upper limits reported by the MEG II~\cite{MEGII:2023ltw}, Belle~\cite{Belle:2021ysv} and BaBar~\cite{BaBar:2009hkt} collaborations. A clear separation is observed between the NO and IO cases. The predicted points lie below the current experimental upper limits, which indicates that the model can satisfy the stringent cLFV constraints while simultaneously accounting for the observed baryon asymmetry of the Universe.

To conclude, through the phenomenological analysis, the model provides a consistent and predictive framework that simultaneously addresses the three main aspects considered in this section, such as the neutrino masses and mixing, the baryon asymmetry of the Universe, and cLFV. The neutrino oscillation data are well reproduced for both normal and inverted orderings, with best-fit points lying within the experimentally allowed regions and with predictions for the effective Majorana mass, the effective electron neutrino mass, and the sum of neutrino masses that fall within the sensitivity of current and next-generation experiments. The resonant leptogenesis mechanism operating at the TeV scale successfully reproduces the observed baryon asymmetry over a broad range of the scaling parameter, with the viable parameter space characterized by heavy neutrino masses in the multi-TeV range and by washout and CP-violation parameters spanning several orders of magnitude. In parallel, the predicted cLFV branching ratios remain well below the current experimental upper limits, ensuring that the model is consistent with existing constraints while providing important consistency checks on the parameter space. Overall, these results demonstrate that the model can accommodate neutrino masses and mixing, TeV-scale leptogenesis, and cLFV constraints.

\section{Summary and outlook}
\label{sec:conclusions}

In this work, an ISS(2,2) model for Majorana neutrinos based on a $D_4$ discrete symmetry has been constructed. The model simultaneously addresses two major puzzles of the SM, namely the origin of neutrino masses and mixing, and the generation of the baryon asymmetry of the Universe. In addition to the two right-handed neutrinos and the two singlet fermions required by the ISS(2,2) mechanism, the model introduces a scalar singlet $S$ and a $D_4$ doublet $\eta$, whose VEVs control the structure of the fermion mass matrices. The $D_4$ and $Z_5$ symmetries play an essential role in forbidding undesired mass terms and in shaping the resulting textures of the charged-lepton and neutrino mass matrices, thereby linking the low-energy neutrino observables to the heavy-neutrino parameters relevant for leptogenesis.

The model provides good theoretical predictions for both NO and IO cases of the neutrino masses. In the NO case, the best-fit point of the atmospheric mixing angle lies in the lower octant, whereas in the IO case it favors the upper octant. But the distributions are broad, there is no clear indication of the octant prediction. The effective Majorana mass $m_{\beta\beta}$ and the effective electron neutrino mass $m_\beta$ are also predicted. For NO, $m_{\beta\beta}$ lies in the range $1$–$4$~meV, well below the sensitivity of current experiments, while for IO it falls in the range $20$–$50$~meV, within reach of KamLAND-Zen~\cite{KamLAND-Zen:2024eml} and of next-generation experiments such as nEXO~\cite{nEXO:2021ujk}, LEGEND-1000~\cite{LEGEND:2021bnm}, and CUPID~\cite{CUPID:2019imh}. All sampled points satisfy the cosmological bound $\sum m_i < 0.12$~eV~\cite{Planck:2018vyg,GAMBITCosmologyWorkgroup:2020rmf}. Long-baseline neutrino experiments such as T2HK~\cite{Hyper-Kamiokande:2018ofw} and NO$\nu$A~\cite{NOvA:2021nfi} will provide improved determinations of the octant of $\theta_{23}$ and perform high-precision measurements of the Dirac CP-violating phase, thereby providing further tests of the model's predictions.

Beyond neutrino oscillation observables, the model yields predictions for cLFV. The branching ratios for the cLFV processes $\mu\to e\gamma$, $\tau\to e\gamma$, and $\tau\to\mu\gamma$ lie many orders of magnitude below the current experimental upper limits reported by MEG II and BaBar, and Belle, ensuring consistency with existing constraints. At the same time, the model predicts a clear separation between the allowed regions and the experimental exclusion contours in the general parameter space, which suggests that the model can satisfy the stringent cLFV constraints while remaining compatible with the observed baryon asymmetry. Given the many orders of magnitude suppression, these rates are unlikely to be accessible even with planned future experiments. Nonetheless, they impose constraints on the parameter space.

The feasibility of thermal leptogenesis within this framework has also been investigated. The numerical results show that, for the lightest pseudo-Dirac pair in the multi-TeV range, the observed baryon asymmetry of the Universe can be successfully reproduced. The viable parameter space corresponds to heavy neutrino masses in the multi-TeV range, with the final asymmetry governed jointly by the CP-violating source and washout effects. The analytic approximations for the efficiency factor, based on the integral and piecewise formulations, reproduce the full numerical solutions of the Boltzmann equations with good accuracy, confirming the reliability of the semi-analytic treatment. The required heavy neutrino masses lie far below the scale of conventional type-I seesaw leptogenesis, and the predicted states may be accessible to future collider searches.

Several directions for future work remain. A more detailed treatment of flavor effects in leptogenesis, including spectator processes and the full set of washout scatterings, would refine the quantitative predictions for the baryon asymmetry. The scalar sector of the model, which plays a key role in generating $D_4$-breaking VEVs and in suppressing unwanted decay channels, could be explored further in connection with collider phenomenology and dark matter. Finally, predictions for neutrinoless double-beta decay may provide complementary tests of the model in the coming years.

\appendix

\section{Tensor-product decompositions of the \texorpdfstring{$D_4$}{D4} group}
\label{app:sol4polyeq}

In this appendix we collect the tensor-product decompositions of the $D_4$ group used in constructing the invariant Lagrangian of section~\ref{sec:model}. The group $D_4$ is the symmetry group of the square, generated by a $\pi/2$ rotation $a$ and a reflection $b$, which satisfy
\begin{equation}
	a^4=e,\qquad b^2=e,\qquad bab=a^{-1}.
\end{equation}
Its irreducible representations consist of four singlets, ${1}_{++}$, ${1}_{+-}$, ${1}_{-+}$, ${1}_{--}$, and one doublet, ${2}$. The labels $s_1,s_2=\pm$ in ${1}_{s_1s_2}$ denote the transformation parities under the generators of the two $Z_2$ subgroups of $D_4$. The Clebsch-Gordan coefficients for the products of these representations are given below, following the conventions of Ref.~\cite{Ishimori:2010au}.

For the product of two doublets,
\begin{equation}
	\begin{aligned}
		\begin{pmatrix}x_1\\ y_1\end{pmatrix}_{{2}}
		\otimes
		\begin{pmatrix}x_2\\ y_2\end{pmatrix}_{{2}}
		={}&
		\left(x_1y_2+x_2y_1\right)_{{1}_{++}}
		\oplus
		\left(x_1y_2-x_2y_1\right)_{{1}_{--}}\\
		&\oplus
		\left(x_1x_2+y_1y_2\right)_{{1}_{+-}}
		\oplus
		\left(x_1x_2-y_1y_2\right)_{{1}_{-+}}.
	\end{aligned}
	\label{eq:doultddecomp}
\end{equation}

For the product of a singlet and a doublet,
\begin{equation}
	\begin{aligned}
		(w)_{{1}_{++}}\otimes
		\begin{pmatrix}x\\ y\end{pmatrix}_{{2}}
		&=
		\begin{pmatrix}wx\\ wy\end{pmatrix}_{{2}},\\[4pt]
		(w)_{{1}_{--}}\otimes
		\begin{pmatrix}x\\ y\end{pmatrix}_{{2}}
		&=
		\begin{pmatrix}wx\\ -wy\end{pmatrix}_{{2}},\\[4pt]
		(w)_{{1}_{+-}}\otimes
		\begin{pmatrix}x\\ y\end{pmatrix}_{{2}}
		&=
		\begin{pmatrix}wy\\ wx\end{pmatrix}_{{2}},\\[4pt]
		(w)_{{1}_{-+}}\otimes
		\begin{pmatrix}x\\ y\end{pmatrix}_{{2}}
		&=
		\begin{pmatrix}wy\\ -wx\end{pmatrix}_{{2}}.
	\end{aligned}
	\label{eq:doultdsingltdecomp}
\end{equation}

For the product of two singlets,
\begin{equation}
	{1}_{s_1s_2}\otimes {1}_{s_1's_2'}
	=
	{1}_{s_1''s_2''},
	\qquad
	s_1''=s_1s_1',\quad s_2''=s_2s_2',
	\label{eq:singltsdec}
\end{equation}
where $s_i,s_i'=\pm$.

Here $x_i$, $y_i$, and $w$ denote fields or products of fields transforming in the indicated representations. Eq.~\eqref{eq:doultddecomp} gives the decomposition of the product of two doublets into the four singlets, eq.~\eqref{eq:doultdsingltdecomp} specifies the products of a singlet with a doublet, and eq.~\eqref{eq:singltsdec} gives the multiplication rule for singlets. These rules are used throughout section~\ref{sec:model} to construct the $D_4$-invariant operators and to determine the allowed Yukawa couplings of the model.


\renewcommand{\tabularxcolumn}[1]{m{#1}}
\newcolumntype{L}{>{\raggedright\arraybackslash}X}
\newcolumntype{C}[1]{>{\centering\arraybackslash}m{#1}}

\section{Forbidden operators and the role of the \texorpdfstring{$D_4$}{D4} and \texorpdfstring{$Z_5$}{Z5} symmetries}
\label{app:DZ5forb}

The flavor symmetries $D_4$ and $Z_5$ are essential for the viability of the model. They forbid a large number of operators that would otherwise generate unwanted mass terms, spoil the ISS structure, or reintroduce the specific couplings that the model is designed to avoid. In particular, $D_4$ forbids operators that would mix the $D_4$ singlet and doublet sectors at the renormalizable level, as well as operators built from products of three doublets. For example, the renormalizable singlet-fermion mass term $\overline{\chi_D^c}\chi_D\eta$ is forbidden by $D_4$ symmetry. Similarly, $Z_5$ forbids operators that would generate bare mass terms for the right-handed neutrinos and the singlet fermions, such as $\overline{\chi_D^c}\chi_D S$ and $\overline{N_D^c}N_D S$, as well as the operator $\overline{L_D}\tilde{H}\chi_D$. The complete list of operators forbidden by each symmetry is given in table~\ref{tab:forbidden_yukawa}. This structure ensures that neutrino masses arise only through the ISS mechanism, with the small lepton-number-violating parameter $\mu$ generated exclusively by the dimension-five operator $\overline{\chi_D^c}\chi_D S^2/\Lambda$, and that the scalar sector remains consistent with the phenomenological requirements discussed in section~\ref{sec:model}.
\begin{table}[htbp]
	\centering
	\small
	\setlength{\tabcolsep}{4pt}
	\begin{tabularx}{\textwidth}{@{}L C{1.5cm}@{}}
		\toprule
		\multicolumn{1}{c}{Forbidden terms} & Prevented by \\
		\midrule
		$\overline{L_e}He_D$, $\overline{L_e}He_DS$, $\overline{L_e}He_DS^*$, $\overline{L_D}He_R$, $\overline{L_D}He_D\eta$, $\overline{L_D}He_RS$, $\overline{L_D}He_RS^*$, $\overline{L_e}\tilde{H}N_D$, $\overline{L_e}\tilde{H}N_DS$, $\overline{L_e}\tilde{H}N_DS^*$, $\overline{L_e}\tilde{H}N_D\eta$, $\overline{N^c_D}\chi_D\eta$, $\overline{{\chi}^c_D}\chi_D\eta$, $\overline{N_D^c}N_D\eta$
		& $D_4$ \\
		\midrule
		$\overline{L_D}\tilde{H}N_DS$, $\overline{L_D}\tilde{H}N_DS^*$, $\overline{L_e}\tilde{H}\chi_D\eta$, $\overline{L_D}\tilde{H}\chi_D$, $\overline{L_D}\tilde{H}\chi_DS$, $\overline{L_D}\tilde{H}\chi_DS^*$, $\overline{N_D^c}N_D$, $\overline{N_D^c}N_DS^2$, $\overline{N_D^c}N_DH^{\dagger}H$, $\overline{N_D^c}N_D\eta^2$, $\overline{N_D^c}N_DS^*S$, $\overline{N_D^c}N_DS$, $\overline{N_D^c}N_DS^*$, $\overline{\chi_D^c}\chi_D$, $\overline{\chi_D^c}\chi_DH^{\dagger }H$, $\overline{\chi_D^c}\chi_DS^*S$, $\overline{\chi_D^c}\chi_D\eta^2$, $\overline{\chi_D^c}\chi_DS$, $\overline{\chi_D^c}\chi_DS^*$, $\overline{N_D^c}\chi_DS$, $\overline{N_D^c}\chi_DS^*$, $\overline{N_D^c}\chi_D(S^*)^2$, $\overline{N_D^c}\chi_DS^2$ 
		& $Z_5$ \\
		\bottomrule
	\end{tabularx}
		\caption{Operators forbidden by the $D_4$ and $Z_5$ symmetries. The upper block lists operators eliminated by $D_4$, and the lower block those eliminated by $Z_5$.}
	\label{tab:forbidden_yukawa}
\end{table}

The operators listed in table~\ref{tab:forbidden_yukawa} fall into three categories: (i) those that would generate tree-level Dirac or Majorana mass terms for heavy neutrinos, (ii) those that would induce mixing between the $D_4$ singlet and doublet sectors and thereby spoil the texture of the mass matrices, and (iii) those that would couple the scalar singlet $S$ directly to the SM Higgs or to the charged leptons. Their absence is what allows the model to reproduce the observed neutrino masses and mixing while simultaneously realizing low-scale leptogenesis.

\section{Thermal leptogenesis in the IO scenario}
\label{app:IO}

\begin{figure}[H]  
	\centering
	\includegraphics[width=1.0\textwidth]{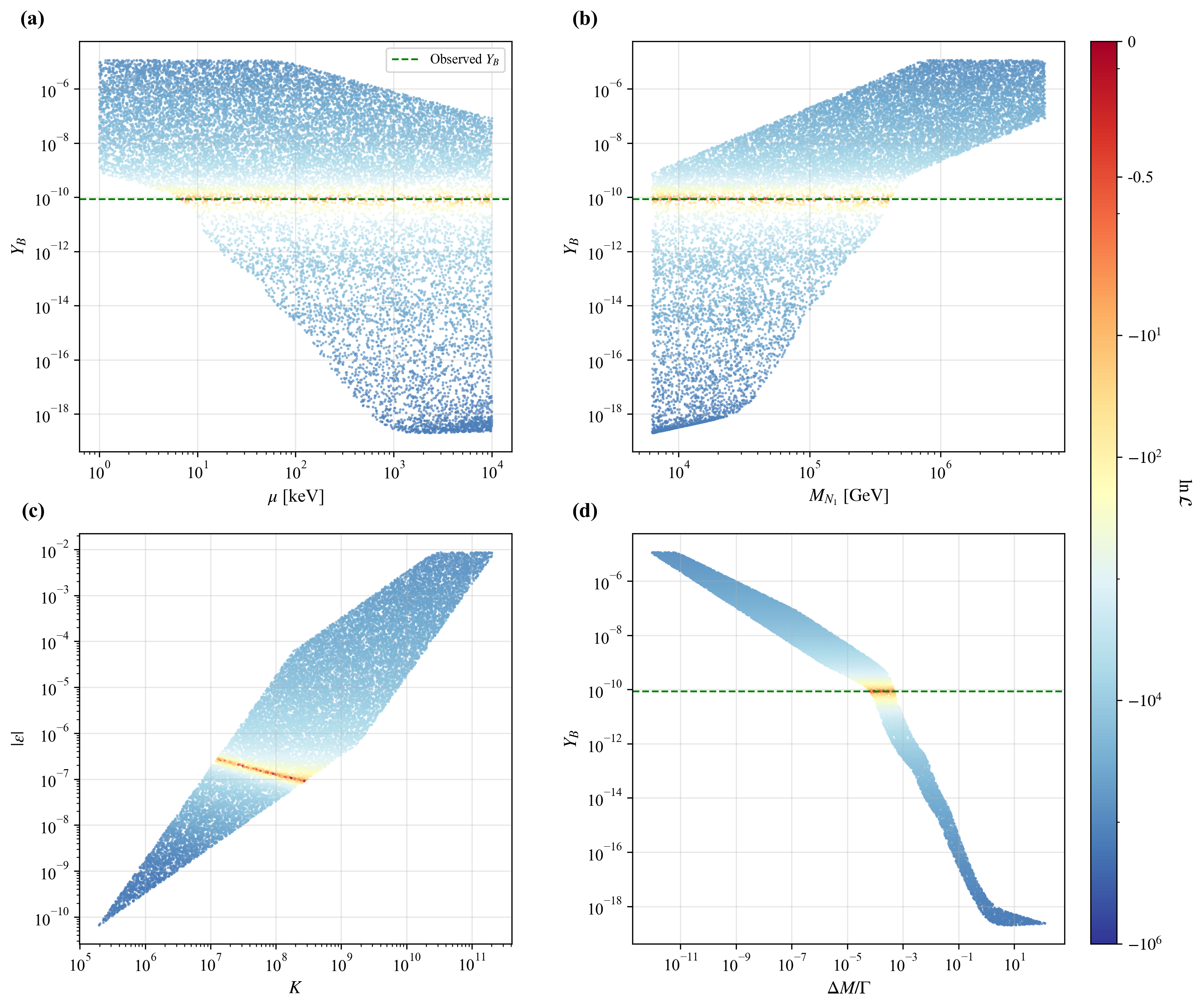}
	\caption{
		Viable parameter space for thermal leptogenesis in the IO scenario. Each panel shows the predicted baryon asymmetry $\eta_B$ (and CP asymmetry) as a function of a different model parameter: (a) the sterile neutrino mass parameter $\mu$; (b) the heavy neutrino mass $M_{N_1}$; (c) the CP asymmetry $\epsilon$ versus the washout parameter $K_i$; and (d) the ratio of the mass splitting to the average decay width, $\Delta M/\Gamma$. The color scale encodes $\ln \mathcal{L}$. The green dashed horizontal line in panels (a), (b), and (d) marks the observed baryon asymmetry $\eta_B^{\rm obs}$.
	}
	\label{fig:app1}
\end{figure}

In section~\ref{subsec:lowsaclleptognss}, we perform a phenomenological analysis of thermal leptogenesis for NO. This appendix collects the corresponding results for the IO case. Since IO and NO possess the same mass‑matrix texture, the structure of the viable parameter space is qualitatively similar. The difference between NO and IO shown in figure~\ref{fig:scheme1_summary} originates from specific directions of parameter tuning. However, this difference is washed out when the three parameters are scanned over their global parameter ranges. Consequently, in figure~\ref{fig:app1}, the leptogenesis results for IO follow a trend similar to that for NO.

As shown in figure~\ref{fig:app1}(a), when the singlet neutrino mass parameter $\mu$ lies in the range $10~\mathrm{keV}$-$10~\mathrm{MeV}$, the model can reproduce the observed baryon asymmetry, and it decreases overall as $\mu$ increases. Figure~\ref{fig:app1}(b) shows that viable leptogenesis can be achieved for pseudo-Dirac pair masses in the range $10$–$100~\mathrm{TeV}$, and the baryon asymmetry increases with the mass $M_{N_1}$. Figure~\ref{fig:app1}(c) displays the correlation between the CP asymmetry $\epsilon$ and the washout parameter $K_i$. The viable region lies roughly in $K_i\sim10^{7}$–$10^{9}$ and $\epsilon\sim10^{-8}$–$10^{-6}$. Figure~\ref{fig:app1}(d) shows that when $\Delta M/\Gamma$ lies between $10^{-4}$ and $10^{-3}$, an asymmetry consistent with the observed value can be obtained, which corresponds to the parameter region where the mass splitting is smaller than the decay width. As $\Delta M/\Gamma$ increases further, the washout effect becomes stronger, causing the generated baryon asymmetry to fall below the observed value. Taken together, these figures confirm that the IO model can successfully reproduce the observed baryon asymmetry, albeit with different preferred ranges for some parameters compared to the NO case.

%
%

\bibliographystyle{RAN}
\bibliography{biblio}

\end{document}